\documentclass[iop]{emulateapj}
\slugcomment{Accepted for publication in ApJ - April 18 2013}
\shorttitle{Young Stars in Nearby Ellipticals}
\shortauthors{Ford \& Bregman}
\begin{document}

\title{Direct Detections of Young Stars in Nearby Elliptical Galaxies\footnotemark[1]}\footnotetext[1]{Based on observations made with the NASA/ESA Hubble Space Telescope, obtained at the Space Telescope Science Institute, which is operated by the Association of Universities for Research in Astronomy, Inc., under NASA contract NAS 5-26555. These observations are associated with program \# 11583.}

\author{H. Alyson Ford\altaffilmark{2,3} and Joel N. Bregman\altaffilmark{2}}
\altaffiltext{2}{Department of Astronomy, University of Michigan, Ann Arbor, MI 48109}
\altaffiltext{3}{National Radio Astronomy Observatory, P.O. Box 2, Green Bank, WV 24944; aford@nrao.edu}

\begin{abstract}
Small amounts of star formation in elliptical galaxies are suggested by several results: surprisingly
young ages from optical line indices, cooling X-ray gas, and mid-IR dust emission. Such
star formation has previously been difficult to directly detect, but using UV Hubble Space Telescope 
(HST) Wide Field Camera 3 (WFC3) imaging, we have identified individual young stars and
star clusters in four nearby ellipticals. This technique is orders of magnitude more 
sensitive than other methods, allowing detections of star formation to $10^{-5}$~M$_\odot$~yr$^{-1}$. 
Ongoing star formation is detected in all galaxies, including three ellipticals that have 
previously exhibited potential signposts of star forming conditions (NGC~4636, NGC~4697, and 
NGC~4374), as well as the typical “red and dead” NGC~3379. The current star 
formation in our closest targets, where we are most complete, is between $1-8\times 10^{-5}$~M$_\odot$~yr$^{-1}$. 
The star formation history was roughly constant from $0.5-1.5$ Gyr (at $3-5\times 10^{-4}$~M$_\odot$~yr$^{-1}$), 
but decreased by a factor of several in the past $0.3$~Gyr. Most star clusters have a mass 
between $10^2 - 10^4$~M$_\odot$. The specific star formation rates of $\sim 10^{-16}$~yr$^{-1}$ 
(at the present day) or $\sim 10^{-14}$~yr$^{-1}$ (when averaging over the past Gyr) imply that 
a fraction $10^{-8}$ of the stellar mass is younger than $100$~Myr and $10^{-5}$ is younger than 
$1$~Gyr, quantifying the level of frosting of recent star formation over the otherwise passive stellar population.
There is no obvious correlation between either the presence or spatial 
distribution of postulated star formation indicators and the star formation we detect.
\end{abstract}

\keywords{Galaxies: elliptical and lenticular, cD --- Galaxies: star clusters: general --- Galaxies: star formation --- 
  Ultraviolet: galaxies --- Ultraviolet: stars}

\section{Introduction}

As a class, normal elliptical galaxies are not actively forming stars.  Their star formation 
rate (SFR) must be significantly less than $1$~M$_\odot$~yr$^{-1}$, or such activity would have been 
identified long ago in the form of prominent H$\beta$ lines or an excess of blue light. However, lower levels
of star formation are suggested by small changes in optical line indices, which
can be used to date stellar populations of galaxies. At constant metallicity, the Balmer line indices 
are sensitive to age, with inferred ages often surprisingly young. For example, in the sample of 
ellipticals in \citet{2000Trager}, 40\% have mean 
ages less than 6 Gyr and several are younger than 3 Gyr.  \citet{2005Denicolo} obtain similar 
results with similar methods. This implies that many elliptical galaxies were forming stars 
quite recently ($z < 0.5$), in conflict with other observations that indicate little star 
formation in ellipticals since $z \approx 1$ \citep{2005Daddi,2005Labbe}. One explanation for 
this discrepancy is the ``frosting'' effect, whereby a small rate of ongoing star formation 
contaminates the Balmer lines, making them relatively strong, while contributing little to the 
mass of the galaxy. The degree of the frosting could be understood if the rate of ongoing 
star formation were known, resulting in more reliable age determinations.

Several mechanisms can lead to small amounts of star formation. One avenue involves stellar 
winds from the ensemble of old stars in a galaxy, amounting to $0.1-1$~M$_\odot$~yr$^{-1}$ for 
$\sim$L$^*$ ellipticals. Most of this stellar mass loss is believed to collide with the ambient 
interstellar medium (ISM) and become heated to X-ray emitting temperatures that are typically 
$5\times 10^{6}$~K \citep{2003Mathews}, although 
the efficiency of this process is unknown and some gas may remain cool \citep{2008Parriott}, 
and occasionally may 
form into new stars. Most of the hot ISM in a galaxy has a cooling time much less than a 
Hubble time, and if not driven out as a galactic wind (through active galactic nuclei or supernovae), 
the gas will radiatively cool.
Radiatively cooling gas is found in $30-40$\% of ellipticals, as apparent from the detection of
the \ion{O}{6} line in Far Ultraviolet Spectroscopic Explorer (FUSE) observations \citep{2005Bregman}, 
implying a cooling rate of $\sim 0.1-0.5$~M$_{\odot}$~yr$^{-1}$. 
Gas emitting \ion{O}{6} is at the peak of the cooling curve, so it will cool to the $10-10^{4}$~K 
range. This cooled gas, confined to the central region ($< 1$~kpc), is a natural source of 
material for star formation. In addition to the internal recycling of galactic gas, there may 
be infall of material onto a galaxy from other smaller galaxies or from ambient group material.

Although the gaseous content of elliptical galaxies is generally dominated by hot 
($5\times 10^{6}$~K) X-ray emitting material with masses of $10^{8}-10^{10}~$M$_\odot$ 
\citep{1991Roberts}, some evidence for cool gas ($< 10^{4}$~K) exists in addition to \ion{O}{6} 
emission. H$\alpha$-emitting material at $\sim 10^{4}$~K is present in the central kpc 
($10-20''$) of most ellipticals \citep{1990Matthews,2000Caon}, although it constitutes a 
relatively small amount of gas ($10^{4}-10^{5}$~M$_\odot$). While few elliptical galaxies have 
detectable amounts of \ion{H}{1} ($<10^{7}-10^{8}$~M$_\odot$) or H$_2$ 
\citep{1991Roberts,2007diSeregoAlighieri}, 5-10\% were detected at 60-100~$\mu$m by IRAS 
\citep{1998Bregman}, showing that dust emission is occasionally present. Extinction by dust 
lanes is also seen in $\sim50$\% of ellipticals, and is generally near the center 
\citep{2005Lauer}. Evidently, cool gas exists in ellipticals and is most common in the 
central region, but we do not know the fate of this material nor whether its presence is 
stable or varies with time.

A good strategy for quantifying the amount of recent star formation in normal ellipticals is to work in the ultraviolet (UV), 
where old, red stars that dominate these galaxies contribute little compared to hot 
horizontal branch stars, post-Asymptotic Giant Branch (p-AGB) stars, and young stars \citep{1999OConnell}. 
Indeed, many near-UV (NUV) observations have been made of early-type galaxies in an attempt to quantify amounts of
recent star formation. 
For instance, using the Wide-Field Planetary Camera 2 (WFPC2) on the
Hubble Space Telescope (HST), \citet{2000Ferreras} found that most ellipticals in a $z=0.4$ cluster 
must have young stellar mass fractions of at least $0.1\%$, and in many cases $1$ to $10\%$, to account for 
their rest frame $2000$\AA\ emission. At lower redshift, \citet{2007Kaviraj} studied a sample of $2100$ 
early-type galaxies selected systematically from the Sloan Digital Sky Survey and found that $30\%$ 
of them had Galaxy Evolution Explorer (GALEX) NUV-optical 
colors that required a young ($<1$~Gyr) component, amounting to $1$--$3\%$ of their stellar mass 
\citep[see also the similar earlier study by][]{2005Yi}.

A handful of UV observations have also been made of individual stars, though these
observations have mostly been limited to the nearest galaxies. Perhaps 
the most comprehensive set comes from \citet{2008Brown}, where the UV population of stars 
in the nearby elliptical M32 (with a distance, $d=800$~kpc) was studied via near and far-UV data from the 
Space Telescope 
Imaging Spectrograph (STIS) on HST \citep[see also][]{2000Brown}
and using stellar evolution models. However, because \citet{2008Brown} focused on determining the UV contribution from
p-AGB stars and other stars at late evolutionary stages, and because the
field of view of STIS was small ($25\times 25\arcsec$), no analysis of young stars in M32 was 
performed. This technique of imaging individual stars has also been applied to nearby lenticular galaxies; for example, 
NGC~5102, a nearby S0 ($d = 4.0$~Mpc), where \citet{1997Deharveng}
used the Faint Object Camera (FOC) on HST to search for individual young stars as point sources
and deduced a limit for the star formation rate of $5\times 10^{-4}$~M$_\odot$~yr$^{-1}$. 
\citet{1997Deharveng} also observed NGC~3115, another nearby S0 ($d = 9.7$ Mpc), though they
detected no stars and concluded that the FOC was not useful for detecting these stars at
the distances of most elliptical galaxies. 
More recently, \citet{2011Crockett} detected clumps of blue stars in the more distant S0 NGC~4150
($d = 13.7$ Mpc), using HST's Wide Field Camera 3 (WFC3), and determined the star formation
history of NGC~4150 by fitting the UV-optical SED on a pixel-by-pixel basis to a two-burst model.

In this paper we present HST WFC3 UV imaging of four nearby elliptical galaxies, which we used to determine 
star formation rates through the direct identification of young stars and star clusters. Star formation rates
were measured to levels of $10^{-5}$~M$_\odot$~yr$^{-1}$, a limit that is much lower than the 
$\la 0.01$~M$_\odot$~yr$^{-1}$ limits that have previously been attained for galaxies
at similar distances \citep{1999OConnell}. 
Our sample, which we describe in \S \ref{sec:targets}, includes three ellipticals 
with potential signposts of star forming conditions, plus one control galaxy with no prior
indication of star formation. 
Observations and photometry are also described in \S\ref{sec:data}. 
We present the detected UV bright sources in \S\ref{sec:population}, and discuss their physical nature in 
\S\ref{sec:ids}. Star formation rates and histories from the UV bright clusters 
are presented in \S\ref{sec:sfr}, and we conclude with a discussion in \S\ref{sec:discussion}.

\section{Data}
\label{sec:data}

\subsection{Sample Selection}
\label{sec:targets}

Our sample consists of nearby normal $\sim$L$^*$ elliptical galaxies, which are located between $10.6$~Mpc 
and the Virgo Cluster, with the farthest targeted galaxy at $d=18.4$~Mpc \citep[all distances were 
determined via surface brightness fluctuations by][]{2001Tonry}. Three normal elliptical galaxies that 
demonstrated potential signposts of star formation were included, as well as one typical elliptical 
that has shown no indication of having current star formation. The observed targets are summarized 
in Table \ref{tab:targets} and are presented below:

\begin{deluxetable*}{lrcccl}
  \tablecaption{Target Summary\label{tab:targets}}
  \tablehead{\colhead{Galaxy} & \colhead{Type} & \colhead{$m-M$} & \colhead{$d$} & \colhead{$R_{\mathrm{eff}}$} & \colhead{Notes}\\
    \colhead{} & \colhead{}  & \colhead{}  & \colhead{(Mpc)} & \colhead{(arcsec)} & \colhead{} }
  \startdata
  NGC~3379 & E1 & 30.12 & 10.6 & 29.9 &  no prior star formation indicator\\
  NGC~4697 & E6 & 30.35 & 11.8 & 42.4 & PAH emission\\
  NGC~4636 & E0-1 & 30.83 & 14.7 & 59.3 & \ion{O}{6} emission, cooling flows\\
  NGC~4374 & E1 & 31.32 & 18.4 & 34.8 & \ion{O}{6} emission, AGN activity
  \enddata
  \tablecomments{Galaxy types are from RC3 \citep{1991deVaucouleurs}, distances are from \citet{2001Tonry},
    and effective radii are from The 2MASS Large Galaxy Atlas \citep[LGA;][]{2003Jarrett}.}
\end{deluxetable*}

NGC~3379 --- We selected the nearest normal elliptical that is old \citep[$\approx 9$~Gyr;][]{2000Trager} 
and has no indication of recent star formation, NGC~3379 (M105). It is the closest galaxy in our sample at 
$d=10.6$~Mpc, and was chosen as the control galaxy to allow tests of whether the possible star formation 
indicators in the other galaxies within our sample, such as \ion{O}{6} emission,
AGN activity, and PAH emission, are reliable tracers of ongoing star formation.

NGC~4697 --- With a distance of $11.8$~Mpc, NGC~4697 is our second closest target and is usually
designated a typical elliptical galaxy. However, from a sample of 30 nearby normal ellipticals observed
with the Spitzer Space Telescope \citep{2006Bregman_a,2006Bregman_b}, it is the only galaxy that exhibited a
polycyclic aromatic hydrocarbon (PAH) feature. Although PAH emission can be produced without star
formation, e.g., via the winds of intermediate-age stellar populations \citep{2010Vega}, it is often
an indicator of such activity.

NGC~4636 --- A cooling flow elliptical within the Virgo Cluster, NGC~4636 has a distance of $14.7$~Mpc. It
has abundant X-ray emitting
gas that cools on short timescales, and exhibits \ion{O}{6} emission \citep{2001Bregman,2005Bregman}. This \ion{O}{6}
emission is an excellent tracer of cooler gas that could feed star formation and indicates a cooling rate of
$0.3$~M$_\odot$~yr$^{-1}$.

NGC~4374 --- The farthest of our targets, NGC~4374 (M84) is in the Virgo Cluster and has a distance of $18.4$~Mpc. 
\ion{O}{6} emission has been detected in this galaxy \citep{2005Bregman}, suggesting the presence of
cooling gas that can feed star formation. Radio jets and AGN activity are also present, which may correlate
with star formation \citep[e.g., see][]{2004Heckman}.

\subsection{Observations and Data Reduction}
\label{sec:observations}

Observations were conducted using HST's WFC3 between 2009 November 23
and 2010 March 29 (Cycle 17, PID 11583). These observations were made using UVIS 
through two filters, F225W and F336W, which are spaced far enough in wavelength that stars
on the  main sequence would stand out in color from possible contaminants such as UV-bright
globular clusters on a color-magnitude diagram (CMD). Not only are the background 
levels in the UV substantially lower than at optical wavelengths, but fluctuations in the 
background due to horizontal branch stars \citep{1993Worthey} are much fainter than young stars,
which are expected to emit strongly in the UV.

Each galaxy was observed over two orbits for each of the F225W and F336W filters.
Dithered observations were performed over each two-orbit sequence, using the UVIS box dithering
pattern with four pointings per orbit to aid in the removal of artifacts. Each pointing had an exposure
time of 600s, for a total of 4800s per filter per galaxy. As the exposure times were the same for all
four of our targets, the data of NGC~3379 and NGC~4697, our closest targets,
are deeper than those of NGC~4636 and NGC~4374.

Data were reduced on-the-fly during retrieval from the Multimission Archive at STScI (MAST), resulting in
pipeline-calibrated and flat-fielded individual exposures. We then removed cosmic rays, 
corrected geometric distortions, and combined the dithered pointings from each orbit using
Multidrizzle \citep{2002Koekemoer}. The resulting drizzled images were used as the reference frame while the
photometry was performed.

\subsection{Photometry}
\label{sec:photometry}

We performed PSF photometry using {\sc dolphot} \citep[version 2.0;][]{2000Dolphin}, a stellar 
photometry package with specific HST instrument modules. In addition to its customized settings for 
WFC3 data, {\sc dolphot} can perform localized background subtraction during the 
photometering, which was crucial due to the background emission near the center of each galaxy that 
needed to be subtracted carefully, particularly in F336W.

For each galaxy, we used the deep multidrizzled F225W image as the reference frame. Initial estimates 
for the position shifts, magnification and rotation were made using the {\it wfc3fitdistort} task, and 
the individual 
dithered exposures were then photometered on a chip by chip basis. All recommended settings for WFC3 data
were used, with the exception of the sky-fitting setting (FitSky). Instead of fitting the sky prior to each photometry 
measurement, the sky was fit inside the PSF region of 10 pixels but outside the photometry aperture of 4 
pixels. This was done to properly take into the account the varying background levels near the center of 
each galaxy, particularly at radii dominated by background levels in F336W. 
The output magnitudes from {\sc dolphot} are automatically calibrated using Vega zeropoints, while
aperture corrections were determined manually using the {\it qphot} task in IRAF.
Reddening corrections are based on the dust maps of \citet{1998Schlegel} using 
$R_V=3.1$\footnote{http://wwwmacho.mcmaster.ca/JAVA/Acurve.html}.

\begin{figure}
  \plotone{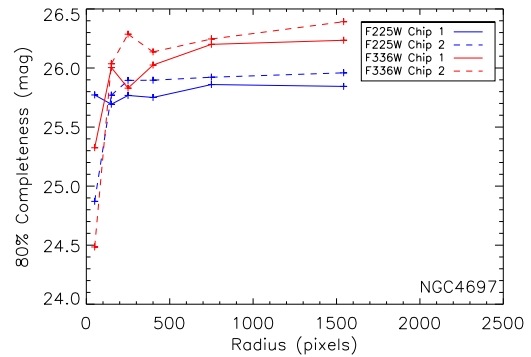}
  \caption{$80\%$ completeness limits as a function of radius for one of our targeted galaxies, NGC~4697, for 
    the F225W (blue) and F336W (red) filters. Chips are distinguished by solid (chip 1) and dashed (chip 2) 
    lines. While there is a radial trend in the ability to detect sources, it is only in the innermost region 
    of the galaxy that the variation is significant.}
  \label{fig:completenessNGC4697}
\end{figure}

Completeness limits were determined based on artificial star tests using approximately $7\times 10^5$
artificial stars. 
We imposed a $1/r$ density distribution for the artificial stars, avoiding the chip gap, so that there 
are approximately equal number of stars in uniformly-spaced radial bins.
Completeness curves were then computed within circular annuli centered on each galaxy. 
We imposed magnitude cuts at the 80\% completeness level in both of the F225W and F336W filters, and then 
derived the accuracy of the recovery based on distance from the center of the galaxy per chip, as there
is a difference in sensitivity per UVIS chip. The 80\% 
completeness limit changes as a function of radius, with sources farther from the center
of the galaxy being easier to detect due to the lower background. 
This trend is demonstrated in Figure \ref{fig:completenessNGC4697}, which is the $80\%$ completeness limit as 
a function of radius for NGC~4697, per filter and chip. The radial trend is weak outside the inner 200 pixels, 
with only a 0.3 magnitude maximum variation, where 200 pixels corresponds to only $3.5\%$ of the area within 
the effective radius of NGC~4697.

After completeness limits were applied, further cuts were made to the detected source list based on 
output from {\sc dolphot}. First, elongated objects, objects that are too sharp, and extended objects
were omitted from the list (i.e., sources with object types greater than 2). All sources containing too 
many bad or saturated pixels were also omitted (i.e., those with error flags greater than 3).
Last, we applied a sharpness cut of 
$|\mathrm{sharpness}_{\mathrm{F225W}}+\mathrm{sharpness}_{\mathrm{F336W}}| \leq 0.4$, which we 
determined via testing to be the most appropriate cut for selecting point sources.

Despite performing the photometry carefully, imposing strict completeness limits based on the
artificial star tests, and applying careful cuts based on {\sc dolphot} quality flags, some contamination from
extended and background sources was apparent in all four galaxy fields as well as all eight blank fields that
were used to estimate background contamination (\S\ref{bcf}). We
visually identified these obvious features, which stood out particularly in F336W, and then removed them by hand. We did not remove anything that 
was ambiguous, i.e., we only removed sources that were clearly elongated and/or clearly part of a nearby background 
galaxy whose bright, star forming regions were detected as individual clumps.
After removing these features, the remaining sources do not appear to correspond to any visible galaxies when 
their positions are overlaid on optical images.

\subsection{Background Control Field}
\label{bcf}

While many background galaxies are easily identifiable by their extended structures, it is certain 
that unresolved background galaxies and point-like AGNs contaminate our source lists. To estimate 
the effect of this contamination in our fields, we searched MAST for WFC3 observations in the 
F225W and F336W filters that had similar exposure times to our data, and used all available data 
that are suitable as blank fields; eight fields from HST PID 11359 (Cycle 17; HUDFs 1-8). We 
retrieved the pipeline-calibrated WFC3 F225W and F336W data for these fields from MAST and 
photometered them identically to our four galaxies. 

Although the blank fields were observed  with the same filters and instrument, the 
exposure times differ from those of our observations: each F225W blank field was observed for 5688s 
(roughly 800s longer than our exposures) and each F336W blank field was observed for 2778s (roughly 
2000s shorter than our exposures). The similar exposure times of the F225W images in the blank fields 
and in our fields, which were used as reference frames in all cases, lends to the photometry being 
performed quite similarly with {\sc dolphot}. As we are particularly interested in blue sources, 
the depth in the redder filter is not as important, though it isn't particularly shallow, being only
0.3 magnitudes more shallow in F336W.

Identical cuts based on output from {\sc dolphot} were applied to the blank fields, including object 
type, error flag, and sharpness. Likewise, all obvious extended sources that were not removed by these 
cuts were removed by hand, as was done in each of our target fields. 
We also imposed a magnitude
cut based on the artificial star tests that were run on our galaxies, to ensure that we did not include
any objects that were only visible due to the deeper F225W data. Aperture and reddening corrections
were applied as described in \S \ref{sec:photometry}. 

We assume all remaining sources within the blank fields to be representative of the unresolved 
background galaxy and point-like AGN population, as there is no reason to expect other contaminants such 
as stars in these fields. We combined the source lists of all eight blank fields and scaled their total 
area to the area of one WFC3 field to estimate the magnitudes and quantity of expected background 
sources in one WFC3 field. While cosmic variance may play a role in the amount of background contamination, 
we cannot account for these effects due to the small area of the sky covered by the eight fields in the HUDF 
and the lack of similar data in other blank fields.

\section{UV-bright Sources}
\label{sec:population}

\begin{figure*}
  \plottwo{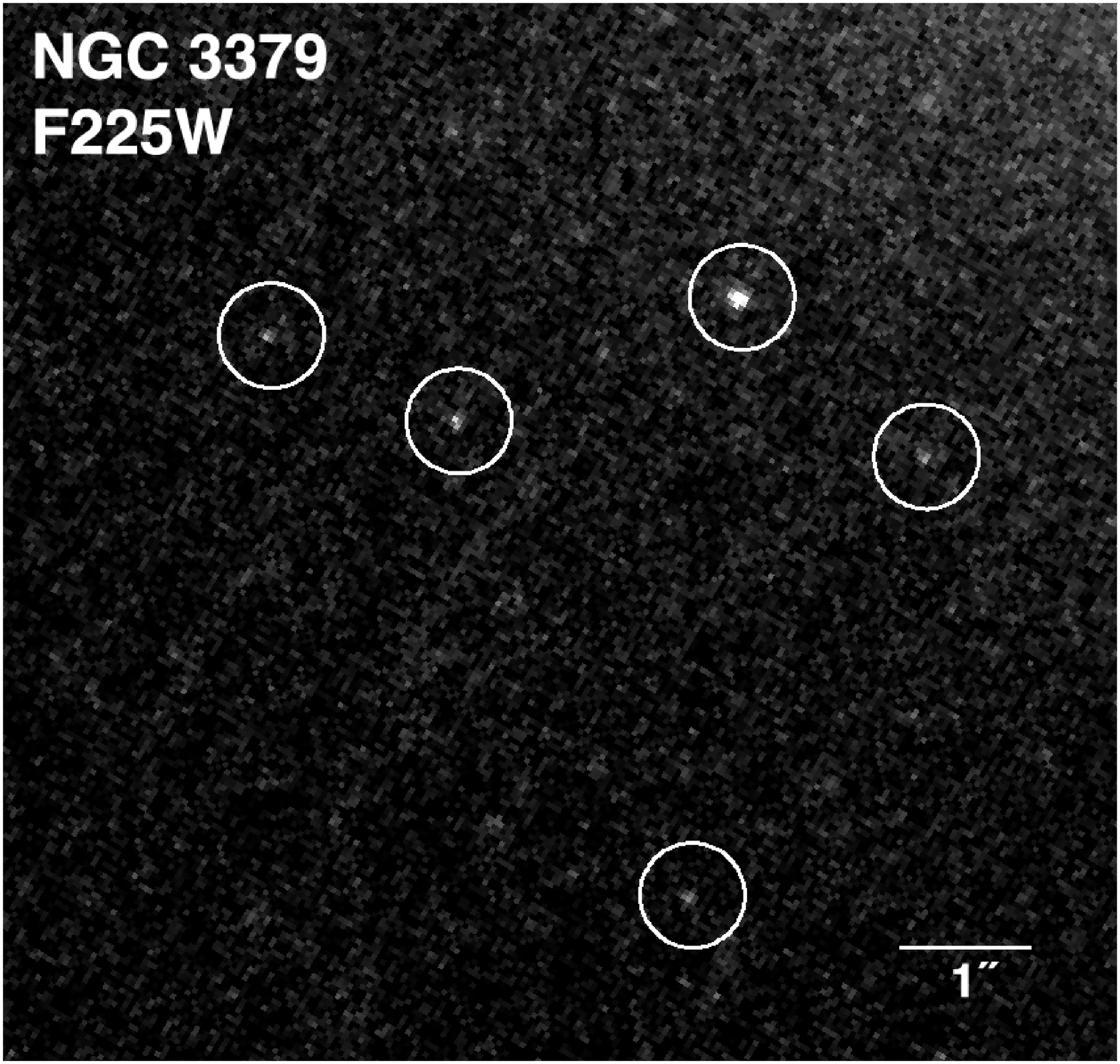}{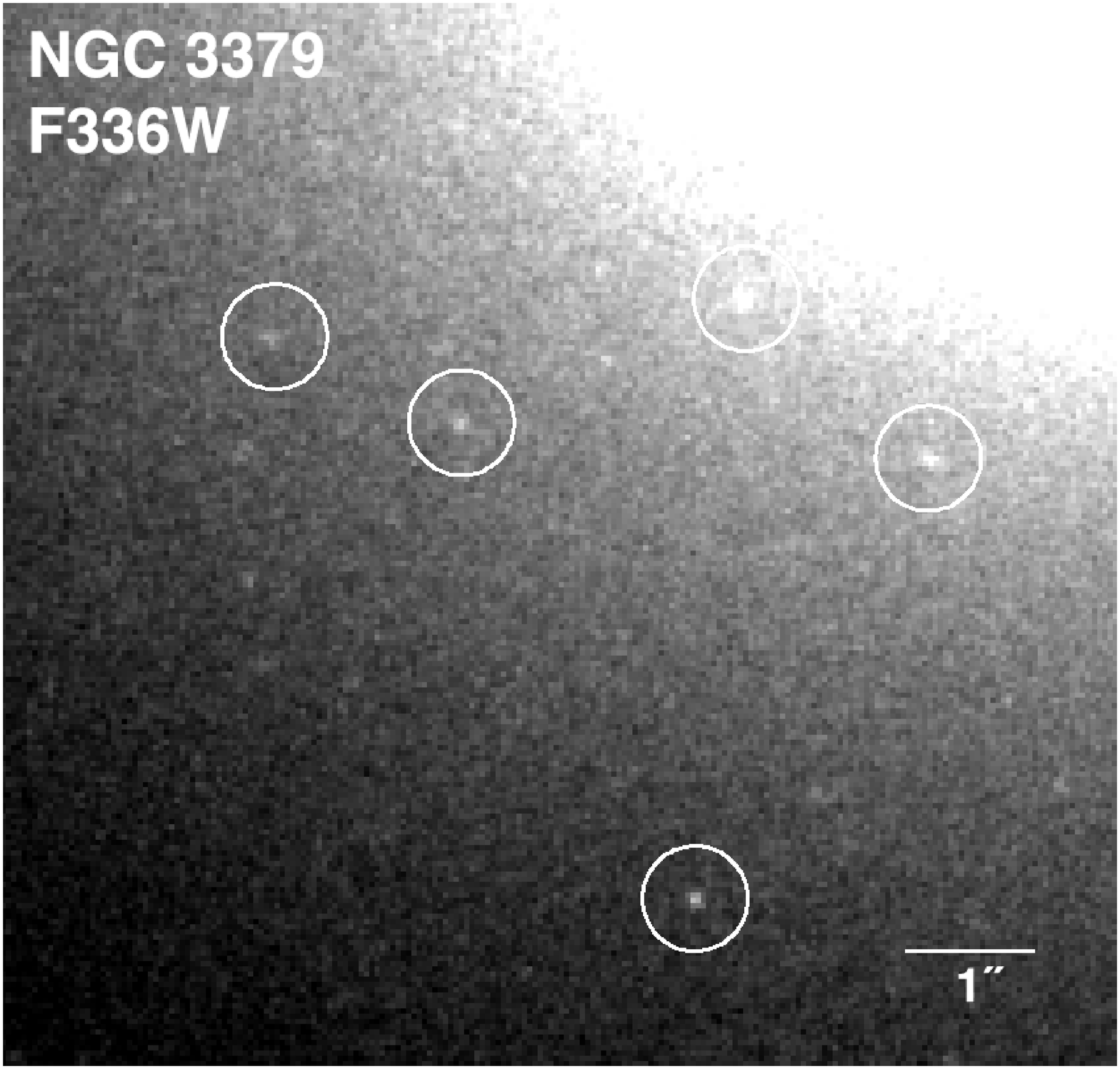}
  \caption{A small portion of the smoothed HST WFC3 F225W (left) and F336W (right) images of NGC~3379. Several 
    UV-bright sources are detected in both filters (circles). For scale, the white, horizontal line in the 
    bottom, right corner of each image indicates $1\arcsec$ (the entire WFC3 field is $162 \times 162\arcsec$).}
  \label{fig:imageszoom}
\end{figure*}

UV-bright sources are detected in all four of our targeted galaxies, though there is variation between the number 
and positions of sources per galaxy. An example of several of these sources can be seen in
Figure~\ref{fig:imageszoom}, which displays a small portion of the observed NGC~3379 field 
in both the F225W and F336W filters. Figure~\ref{fig:imageszoom} also demonstrates a noticeable 
difference in the 
surface brightness of NGC~3379 per filter, where the surface brightness in F225W is extremely 
low, even compared to that of F336W. This demonstrates the importance of removing the background emission
surrounding individual sources carefully during the photometry process, as discussed
in \S \ref{sec:photometry}. It also demonstrates the usefulness of the F225W and F336W filters in searching 
for young stars, as the optical filters would be even more dominated by each galaxy's surface brightness,
rendering this experiment impossible.

\subsection{Color-Magnitude Diagrams}
\label{sec:cmds}

\begin{figure*}
  \plottwo{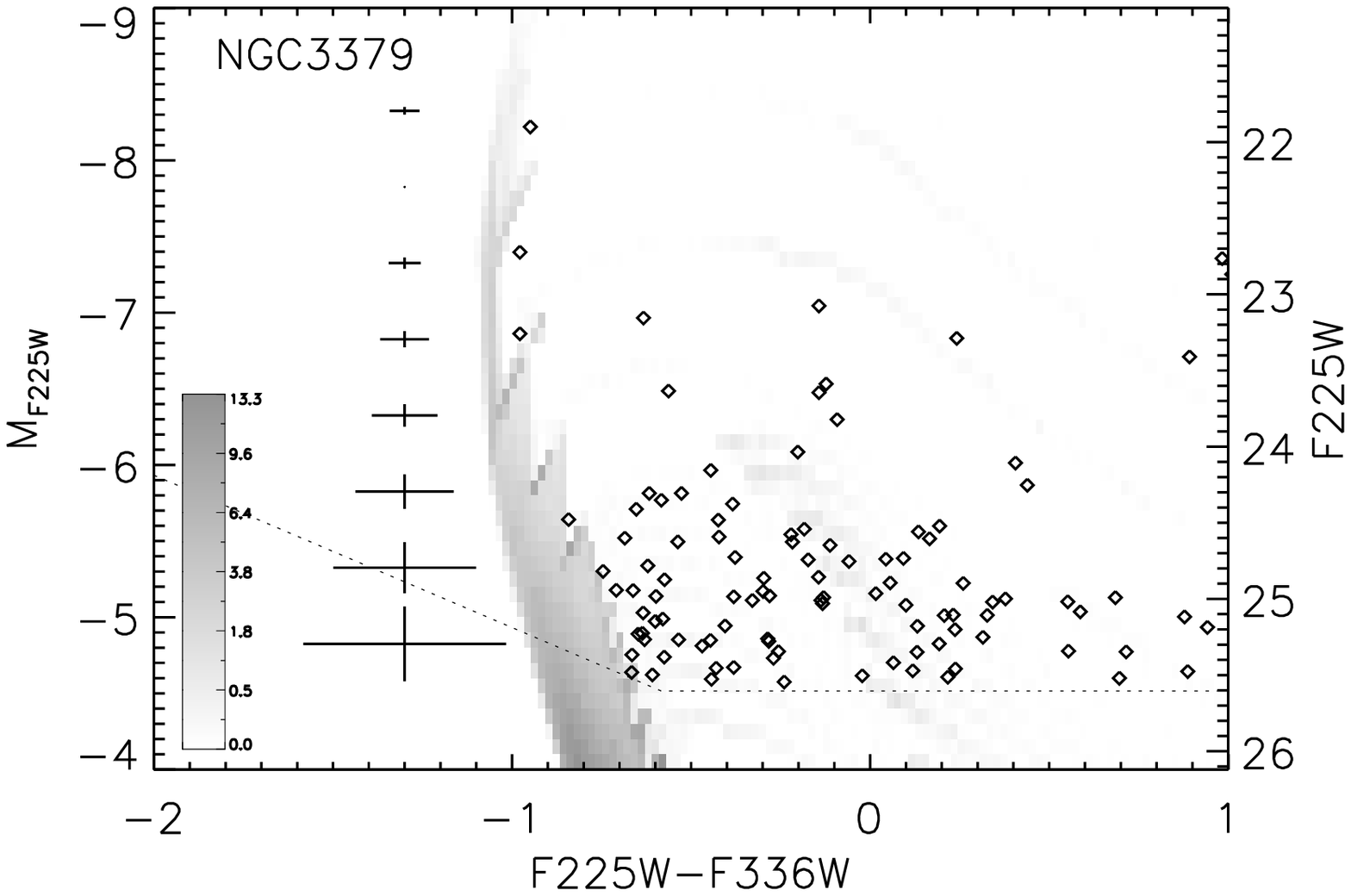}{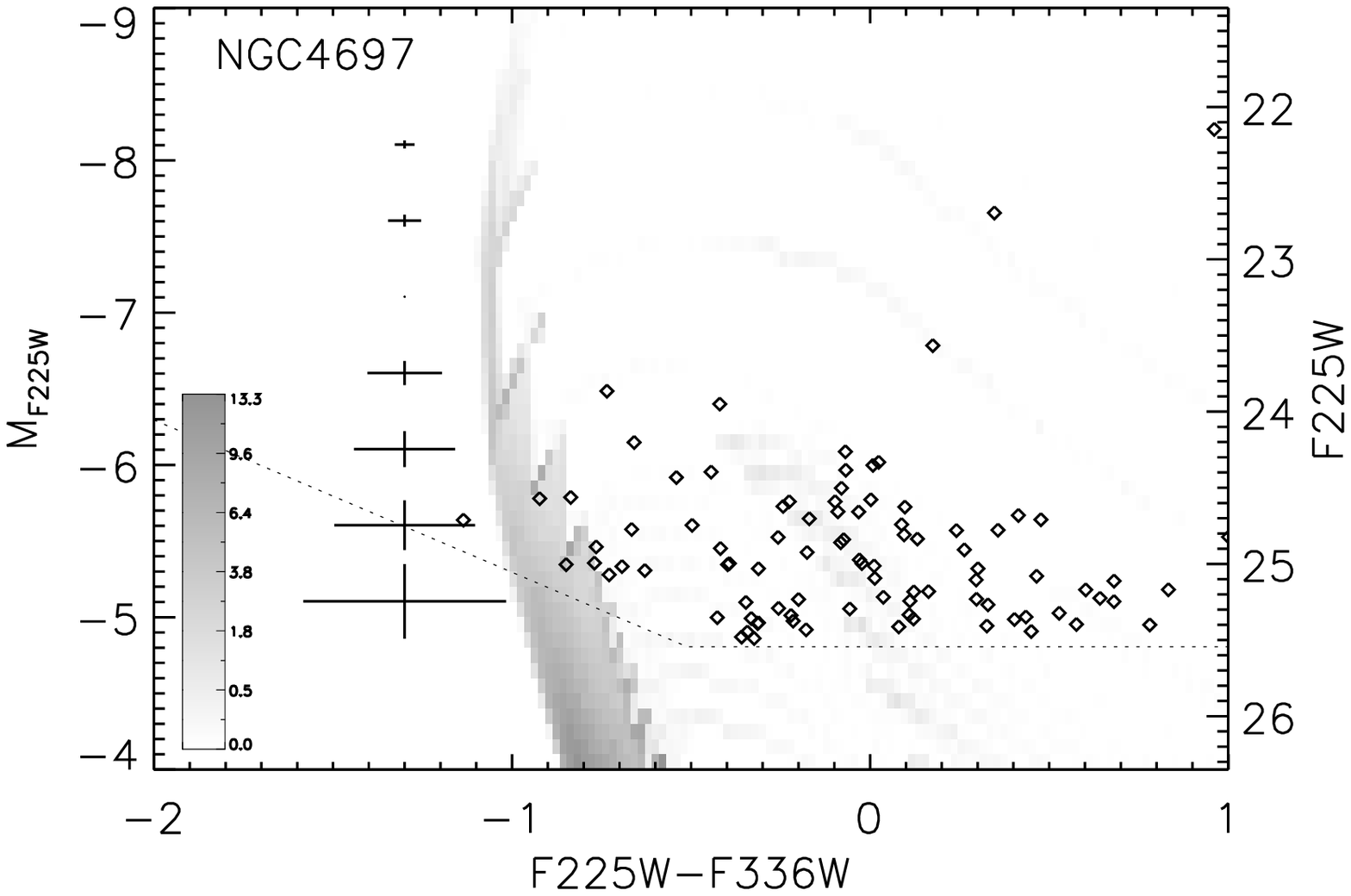}
  \plottwo{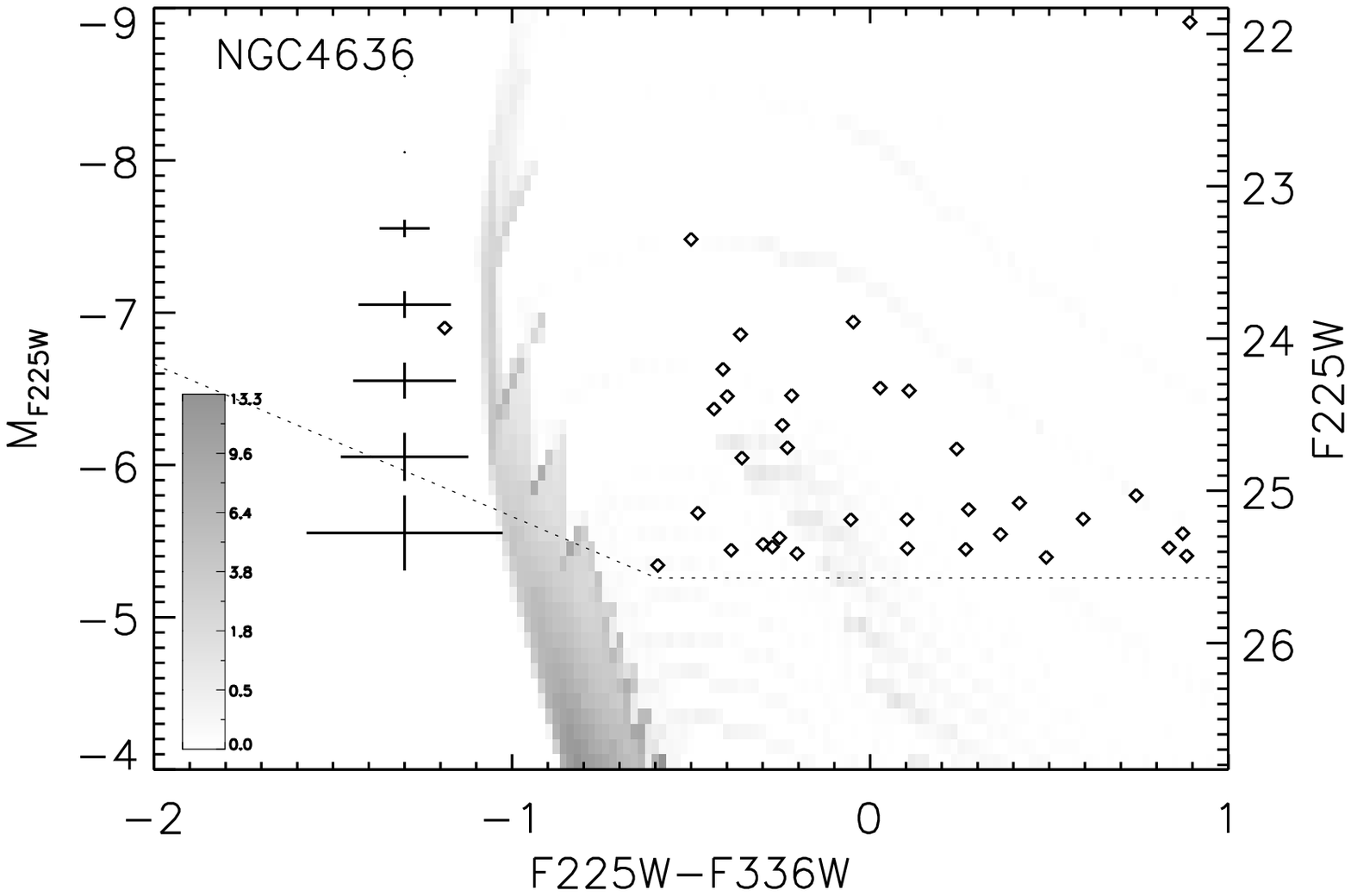}{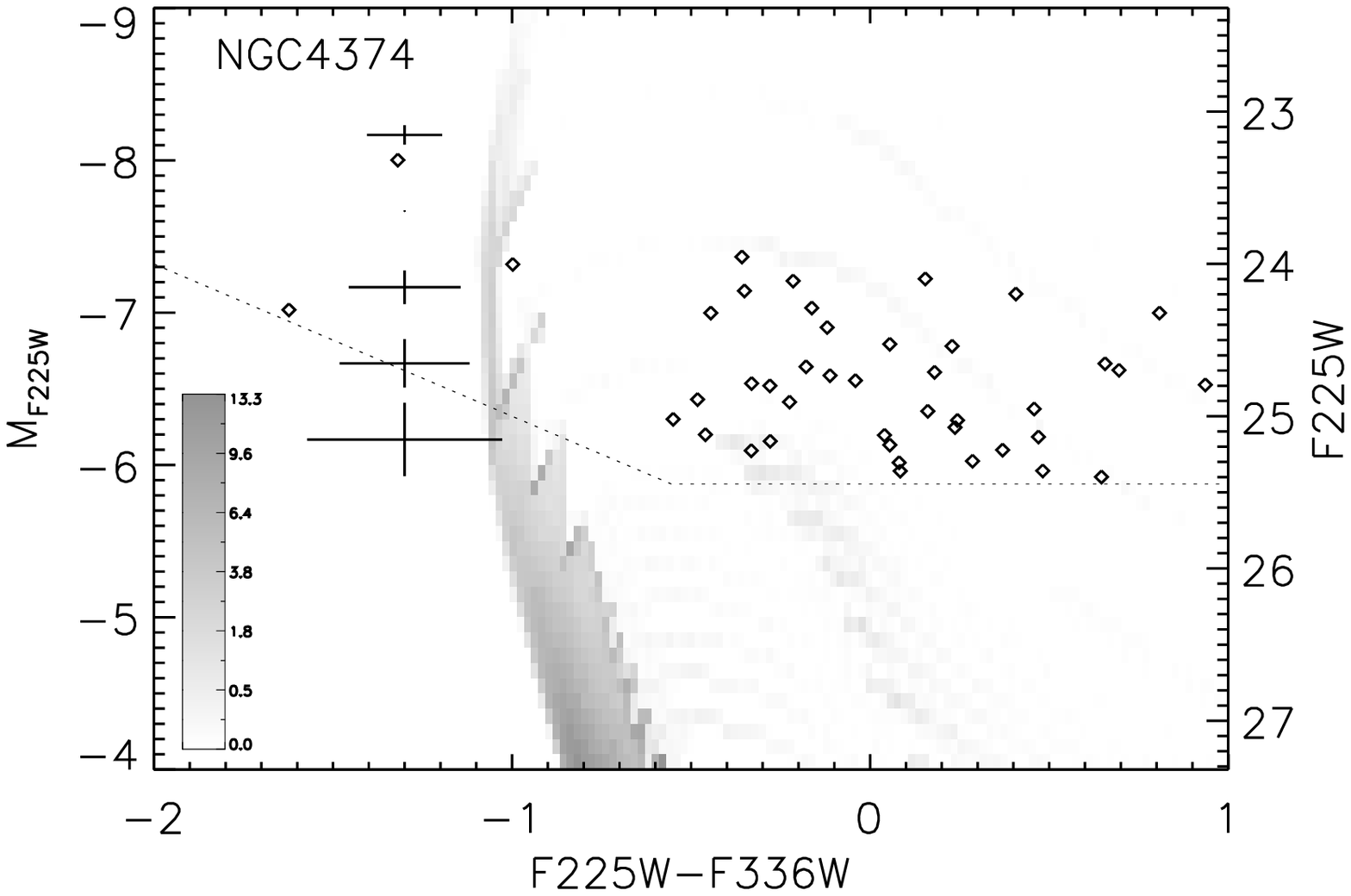}
  \caption{CMDs of targeted galaxies, showing all detected sources with $\mathrm{F225W}-\mathrm{F336W} < 1$
    (diamonds). Apparent magnitude is on the right y-axis while absolute magnitude is on the left y-axis.
    The grayscale Hess diagram indicates the probability of a star being at a given location on the CMD assuming a
    Salpeter IMF and constant SFR over the past 100~Myr, in units of $10^{-6}$ stars per bin of width 0.02 mag and
    height 0.1 mag, and is based on stellar evolutionary tracks from {\it BaSTI}
    \citep{2004Pietrinferni}. The choppy appearance of the Hess diagram is an artifact due to the interpolation between different
    mass tracks. All sources with colors $>1$ are omitted from these CMDs as they are
    predominantly globular clusters, which are beyond the scope of this study. Error bars indicate the
    mean uncertainty, as estimated by {\sc dolphot}, for sources within bins of 0.5 in F225W magnitude.
    Sources are detected in all of our targets, though there are more detected in the nearest targets.
    Dashed lines show the deepest apparent magnitude limits. }
  \label{fig:cmds}
\end{figure*}

Color-magnitude diagrams for all four targeted galaxies are presented in Figure~\ref{fig:cmds}. 
Magnitude limits were determined via artificial star tests, and the deepest magnitude limits are indicated
on the CMDs by dashed lines.
The apparent magnitude limits, which are a function of radius, typically ranged from 
$25.4 < \mathrm{F225W} < 25.7$  and $25.5 < \mathrm{F336W} < 26.3$. These
variations were also dependent on chip, as the UVIS chip 2 is more sensitive than chip 1 by 0.2 magnitudes. 
The inner radial bin, with a radius of 100 pixels, had a much shallower limiting magnitude than
the other radial bins, being about 1 magnitude shallower than the deepest limit, except in NGC~4374 where 
there is a lot of structure in the innermost bin, resulting in the inner bin being about 2 magnitudes
shallower. However, the vast majority of sources are not within the first radial bin as it encloses
very little area ($\sim 0.2\%$) of the entire WFC3 field. The absolute magnitude limit is strongly 
dependent on 
the distance to each of these galaxies, resulting in much deeper data for NGC~3379 and NGC~4697 than for
our other two targets, which accounts for much of the discrepancy in the number of detected sources for 
each galaxy. 

Sources are overlaid on a grayscale Hess diagram, which indicates the probability of stars 
at a given location on the CMD assuming a Salpeter initial mass function (IMF) and a constant star formation rate (SFR) 
over the past 100~Myr. The Hess diagram is based on stellar evolutionary tracks from {\it BaSTI} 
\citep{2004Pietrinferni}, assuming solar metallicity, solar alpha abundance, $\eta=0.4$ canonical models, 
and a Salpeter IMF from 2 to 30 M$_\odot$ (we are limited at the high mass end by the absence of more 
massive {\it BaSTI} tracks, and lower mass stars will not be observable with our data).
The numbers have been rescaled to account for lower-mass stars assuming a \citet{2003Chabrier} IMF.

As globular clusters emit in the UV and are commonly found in ellipticals, we determined the colors of 
globular clusters in the F225W and F336W filters using the existing \citet{2005Dirsch} catalogue for NGC~4636, 
where we matched the positions of optically identified globular clusters to detected sources in our sample. 
Out of $\sim 80$ matched globular clusters, only 1 had a color blueward of $\mathrm{F225W-F336W} < 1$. We 
therefore expect very little, if any, contamination from globular clusters within the color range of 
$\mathrm{F225W-F336W} < 1$. Redward of $\mathrm{F225W-F336W} = 1$, however, the CMDs are dominated by globular 
clusters, and so we limit the CMD color range to $\mathrm{F225W-F336W} < 1$ as the globular cluster 
populations of these galaxies are beyond the scope of this study. 

\subsection{Source Distribution}

\begin{figure*}
  \plottwo{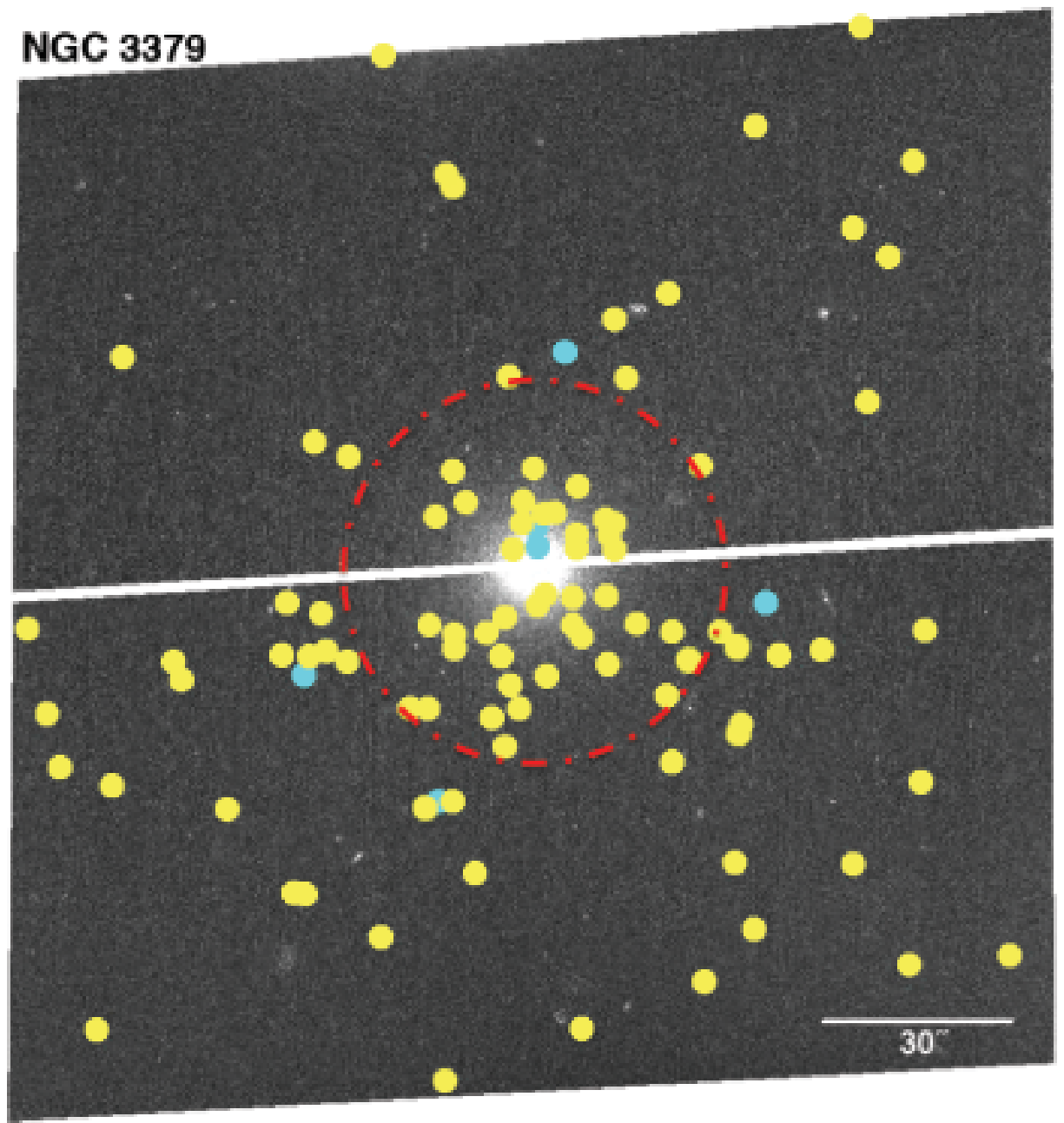}{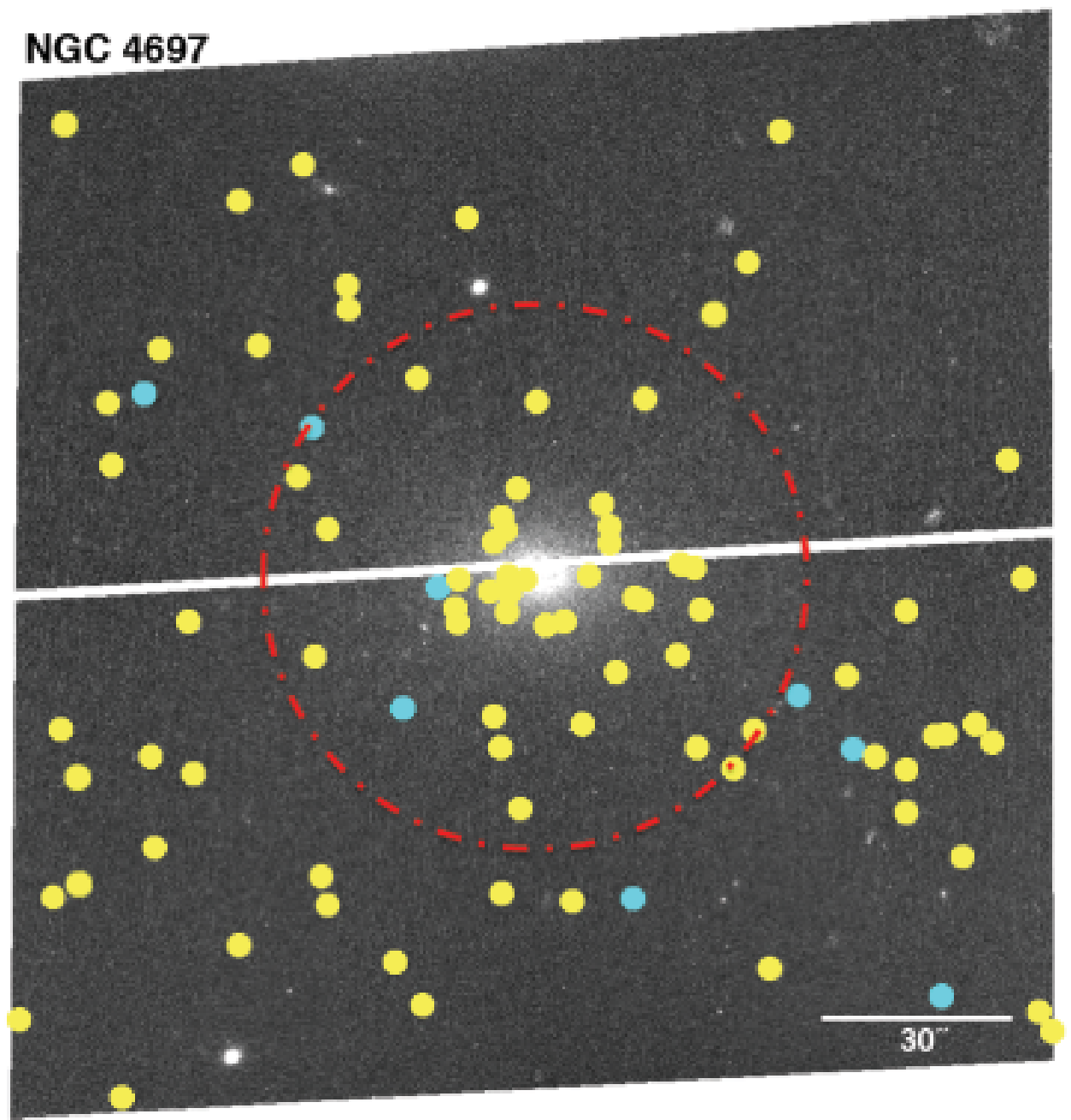}
  \plottwo{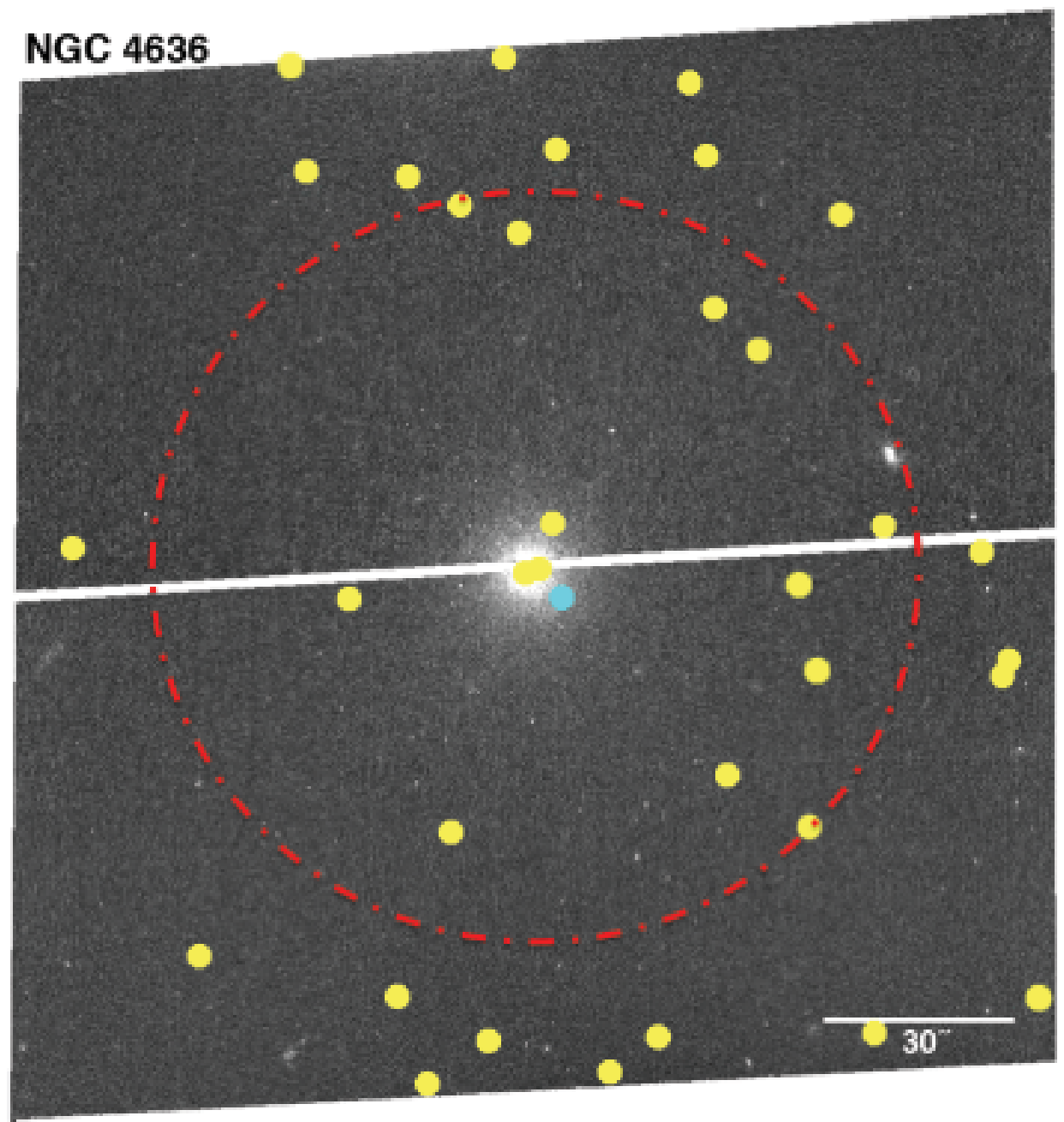}{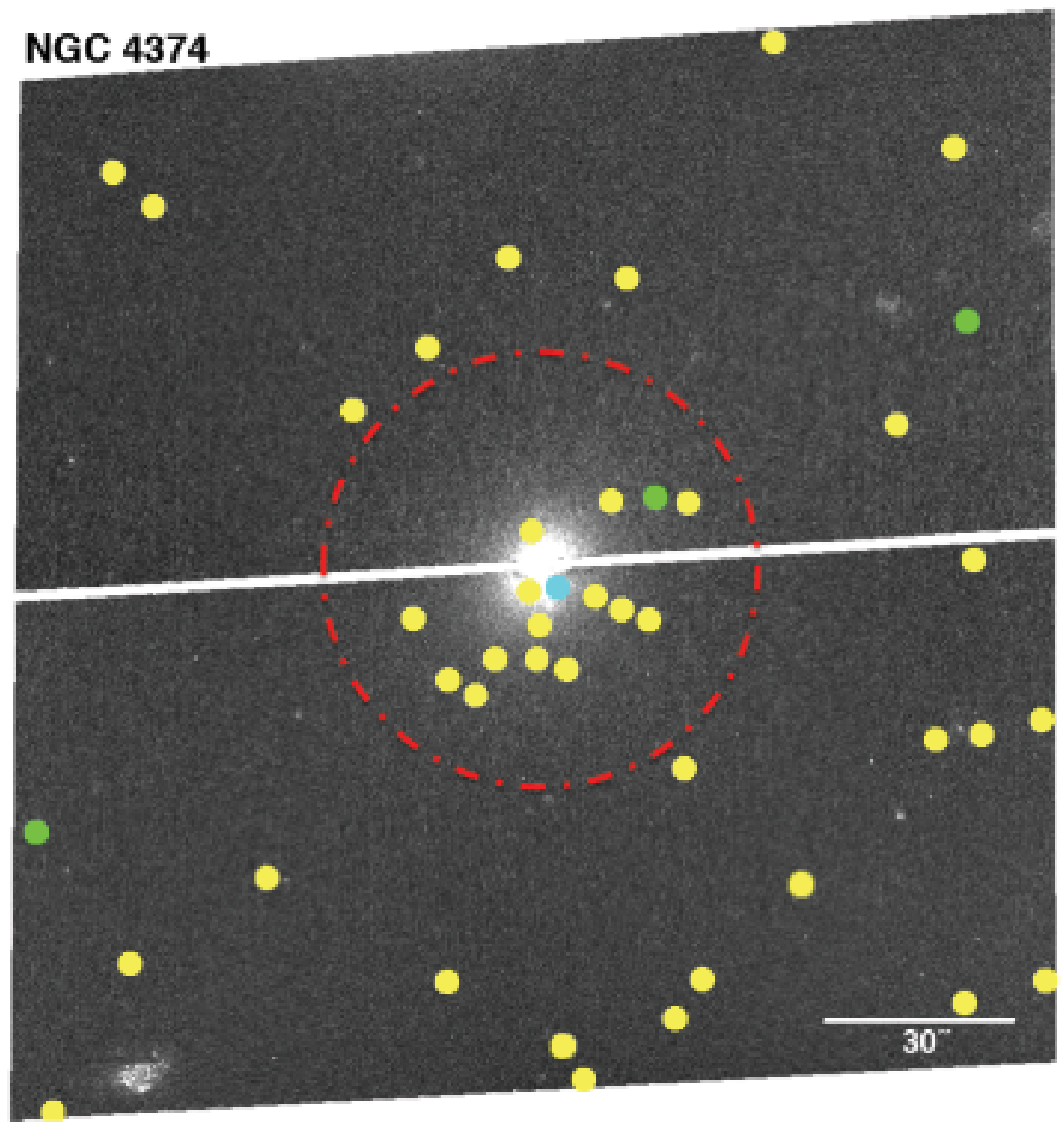}
  \caption{Smoothed HST WFC3 F225W images of observed galaxies. Colored dots represent the  
    positions of UV-bright sources in our sample, and are color-coded by their $\mathrm{F225W-F336W}$ value:
    blue dots represent sources that are near the main sequence ($-1.3 < \mathrm{F225W-F336W} < -0.7$), yellow 
    dots are sources with colors to the red of the main sequence, and green dots are sources to the blue of 
    the main sequence. The dot-dashed red annuli show the effective radius. UV-bright sources are detected in 
    all four galaxies, though there is variation between the number and position of sources per galaxy. For 
    example, many sources are concentrated around the centers of NGC~3379, NGC~4697, and NGC~4374, while there 
    are very few sources near the center of NGC~4636. The total WFC3 field is $162\times 162 \arcsec$. For 
    reference, the white, horizontal line in the bottom, right corner denotes $30~\arcsec$.
    The sources that are visible on these images without overlaid dots did not meet our selection criteria.}
  \label{fig:images}
\end{figure*}

WFC3 F225W images for each galaxy are shown in Figure \ref{fig:images}, where the positions of sources
in our sample (all sources with $\mathrm{F225W-F336W} <1$) are marked by colored dots.
Each color represents a different F225W$-$F336W range: blue represents all sources that are near the
main sequence ($-1.3 < \mathrm{F225W-F336W} < -0.7$), yellow represents all sources with colors to the red of 
the main sequence ($\mathrm{F225W-F336W} > -0.7$), and green represents all sources to the blue of the main sequence 
($\mathrm{F225W-F336W} < -1.3$); there are only 3 green points, all of which are in NGC~4374. 
Many sources are concentrated around the centers of NGC~3379, NGC~4697, and NGC~4374, while there are  
fewer sources near the center of NGC~4636. Also, most sources are not within the color range of the main sequence,
and instead have colors within $-0.7 < \mathrm{F225W-F336W} < 1$.

\subsection{Radial Surface Densities}

\begin{figure*}
  \plottwo{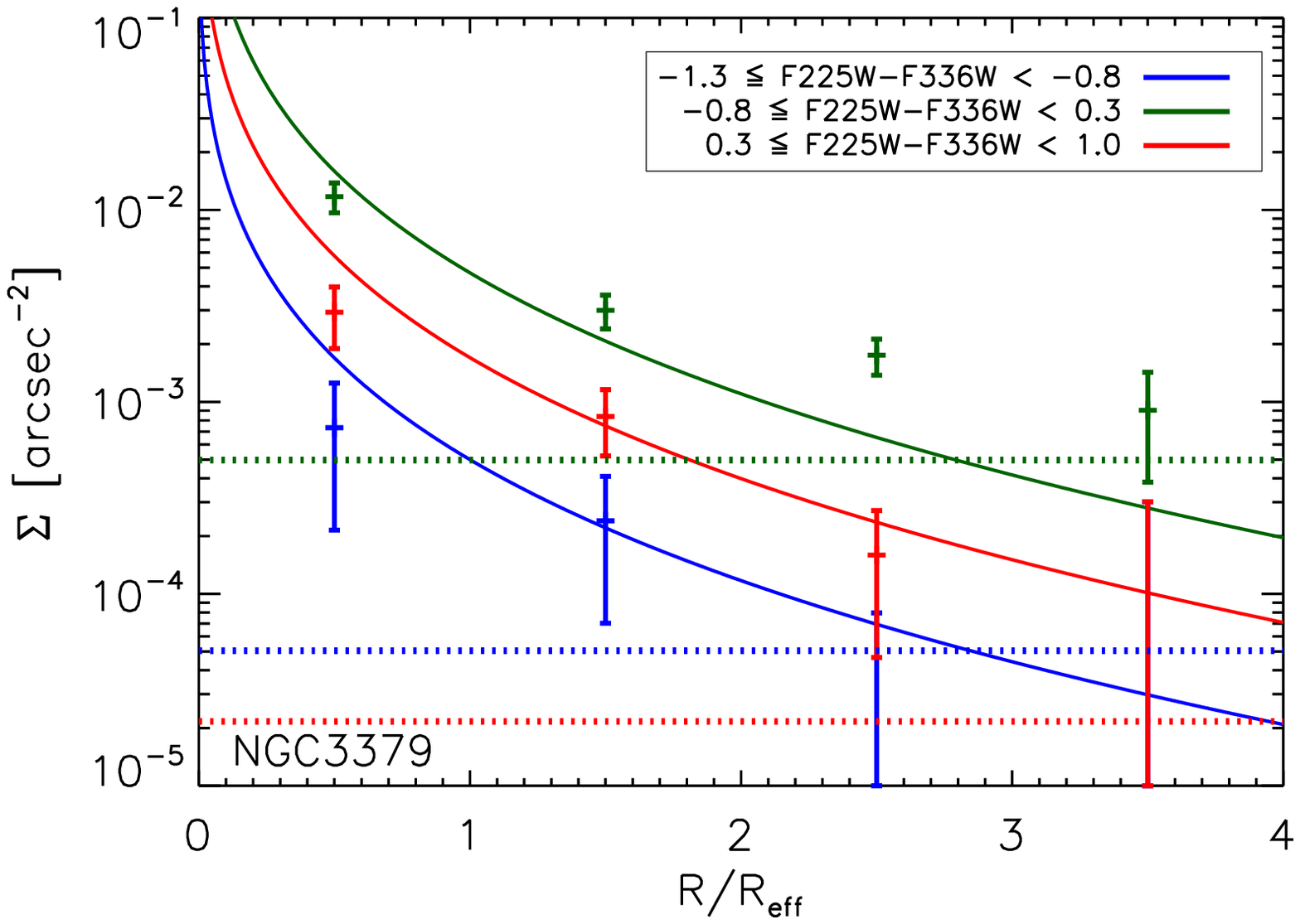}{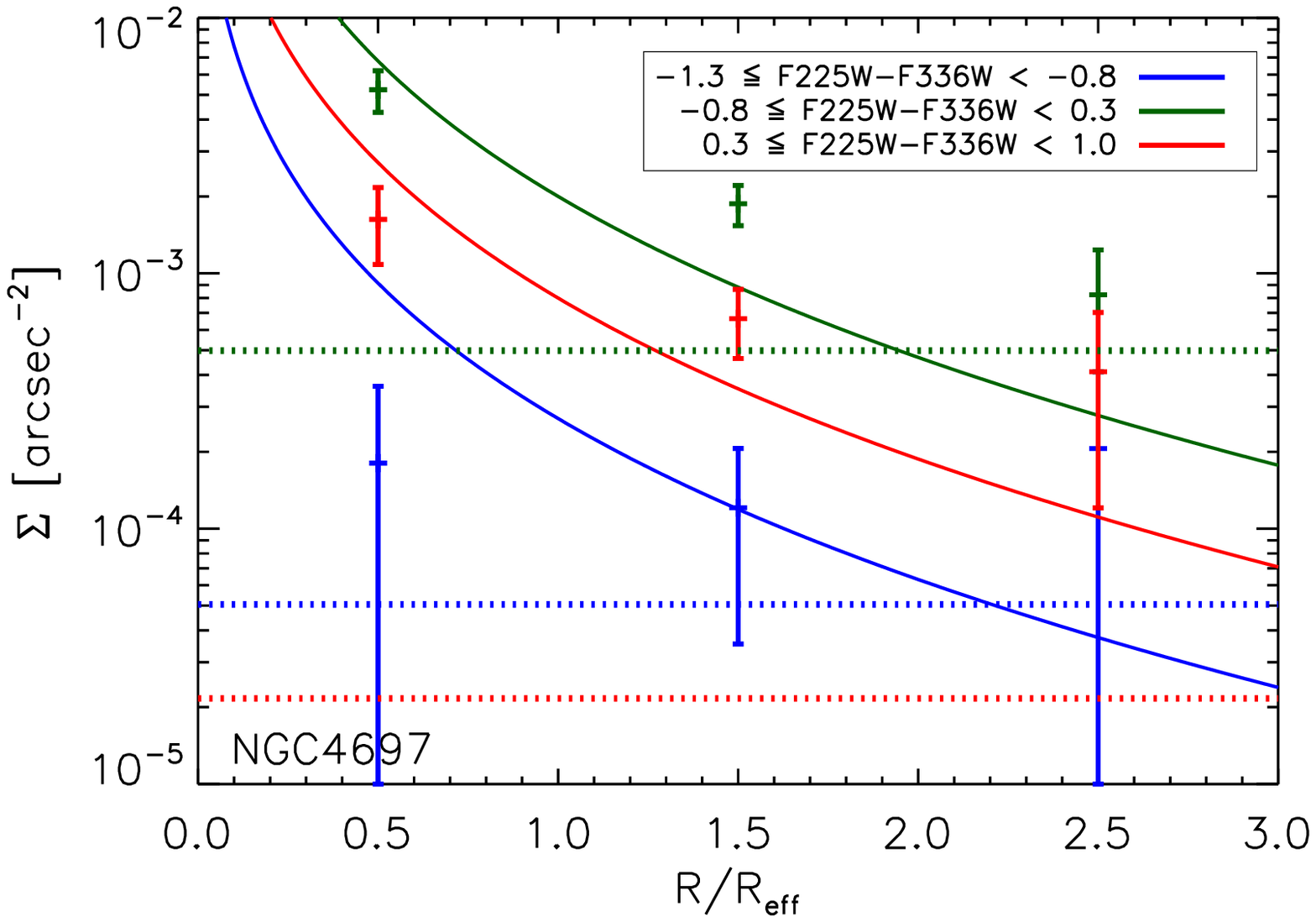}
  \plottwo{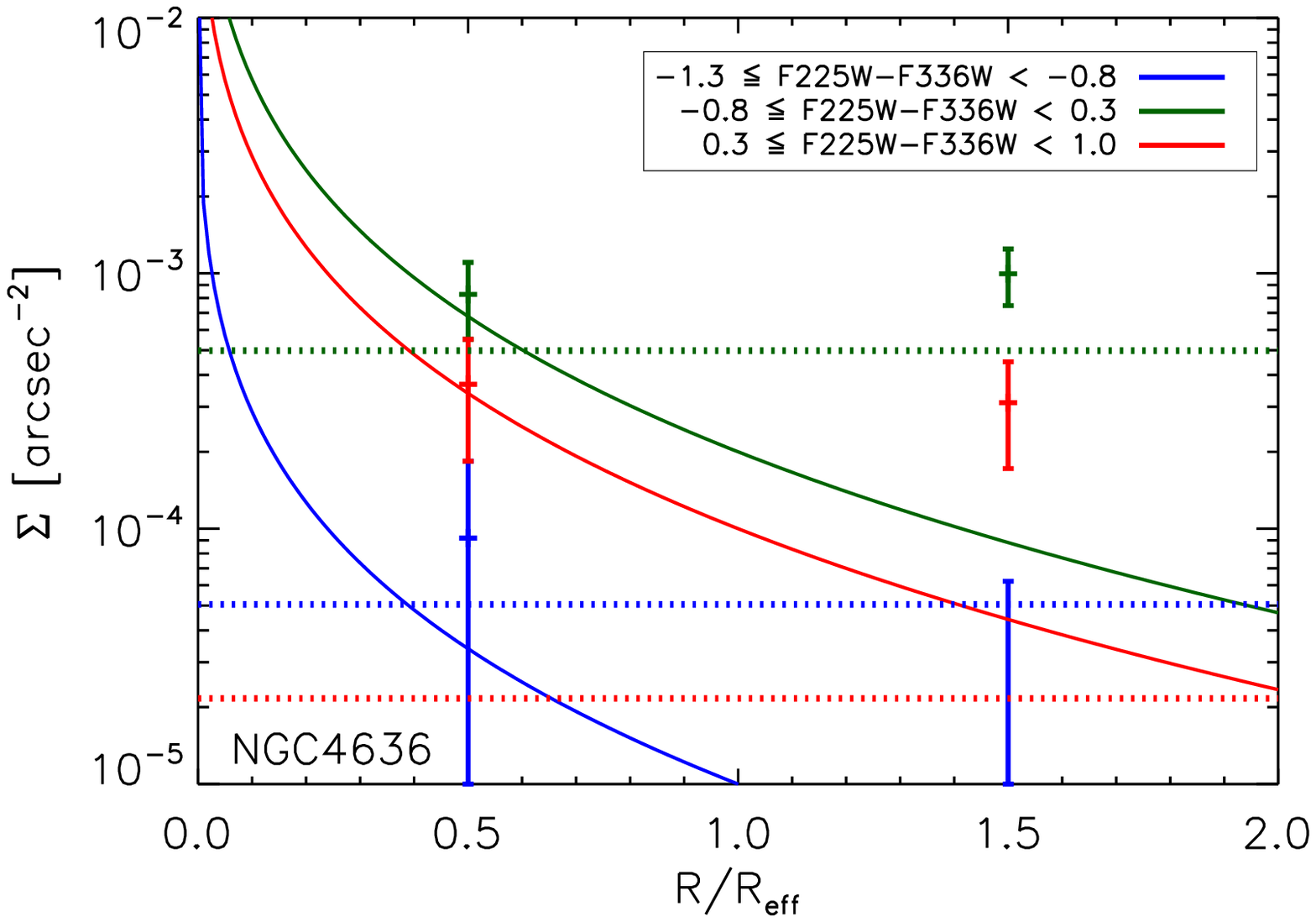}{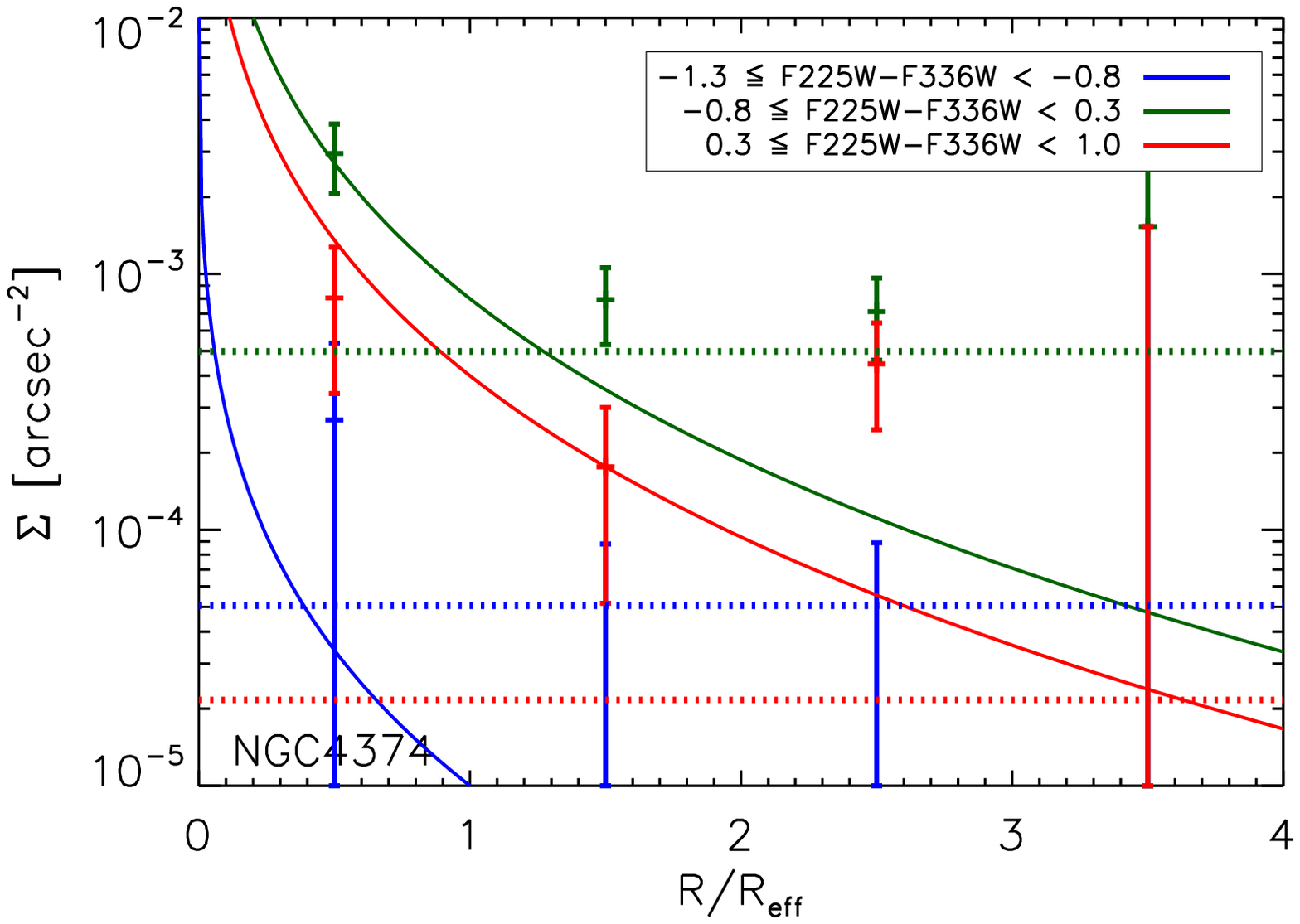}
  \caption{Radial surface density profiles of UV-bright sources within targeted galaxies (points). Horizontal dotted 
    lines are the mean density of background sources and solid curves are de Vaucouleurs profiles 
    \citep{1948deVaucouleurs} arbitrarily scaled in amplitude for reference. Different colors represent 
    different F225W$-$F336W ranges, where blue spans colors expected to be dominated by  main sequence stars 
    ($-1.3 \leq \mathrm{F225W-F336W} < -0.8$), green spans the range expected to be dominated by open clusters and 
    background galaxies ($-0.8 \leq \mathrm{F225W-F336W} < 0.3$), and red spans colors greater than these 
    two ranges ($0.3 \leq \mathrm{F225W-F336W} < 1.0$). Both the red and blue horizontal dotted lines are 
    at very low levels, while the green dotted lines are elevated, showing the background galaxy level. 
  }
  \label{fig:radprofs}
\end{figure*}

\begin{figure}
  \plotone{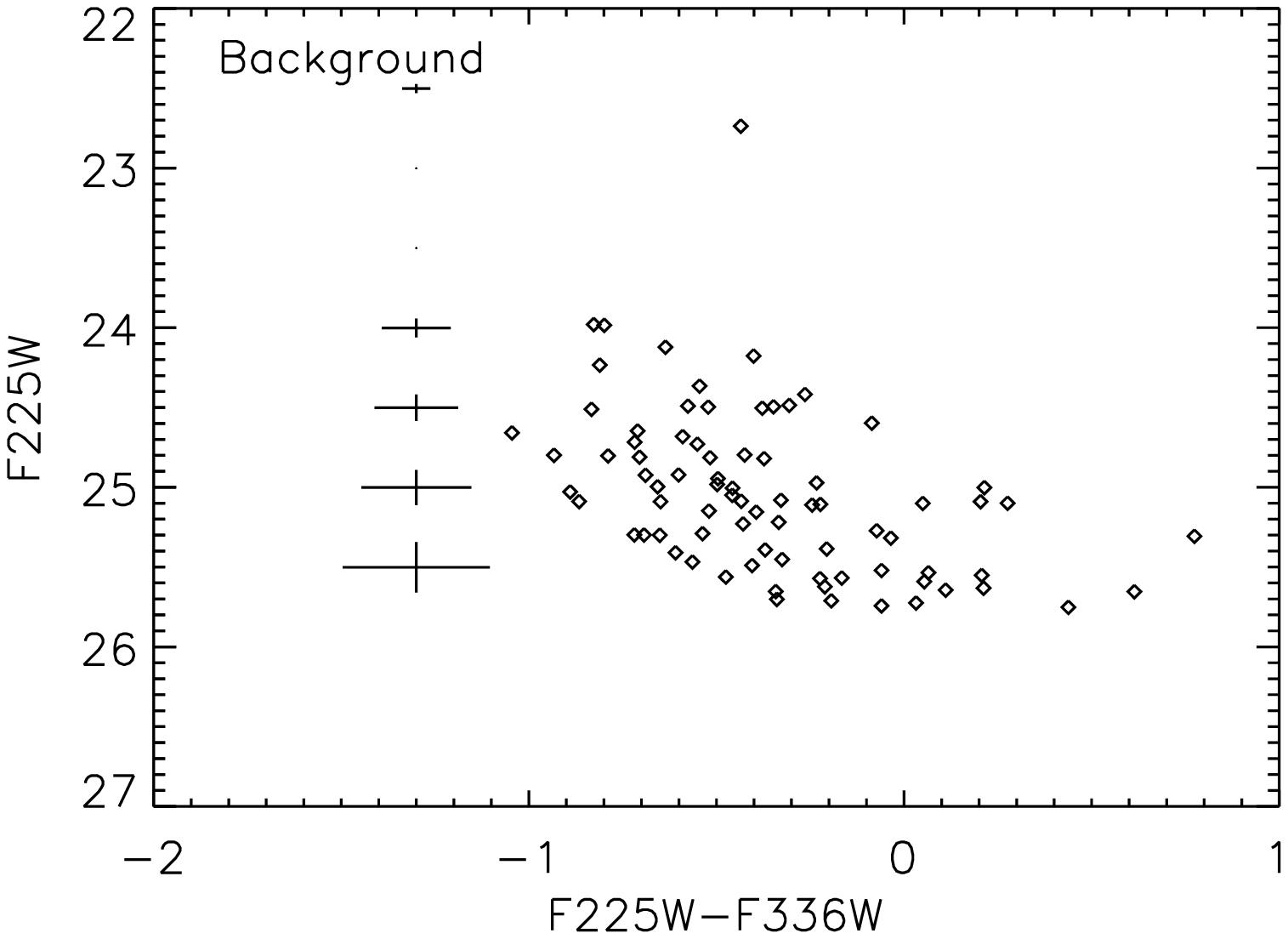}
  \caption{CMD of background control field, determined by combining eight blank HUDF fields.
    All sources are confined to colors between $-1.1$ and $0.8$, with most between $-0.8$ and $0.3$, suggesting that 
    the majority of contamination will be found within this range.}
  \label{fig:cmdhudfs}
\end{figure}

Radial surface density profiles of the detected sources are given as a function of effective radius in 
Figure~\ref{fig:radprofs}. The dotted, horizontal lines in each plot represent the mean density of 
background sources determined from the background control field, the CMD for which is shown in Figure~\ref{fig:cmdhudfs}
and was determined by combining the eight blank HUDF fields. This background has
not been subtracted from the target galaxies CMDs in Figure~\ref{fig:cmds}.
The solid, curved lines represent de Vaucouleurs profiles \citep{1948deVaucouleurs} that are scaled by each galaxy's 
corresponding $R_{\mathrm{eff}}$ and are arbitrarily scaled in amplitude for reference. Different F225W$-$F336W 
ranges are represented by different colors, where blue represents sources that have colors expected of individual 
young stars based on stellar evolutionary tracks, green spans colors expected of background galaxies 
and AGNs (see Figure~\ref{fig:cmdhudfs}), as well as other potential objects that are off the main sequence, and red represents the 
remaining sources with F225W$-$F336W redder than that expected from the background sample and main 
sequence stars. The surface densities account for unobservable regions within each radial bin, such as the corners of 
the fields and the chip gap.

The majority of sources are in the intermediate (``background'') color range. In both NGC~3379 and NGC~4697, the 
surface density of sources in all three color ranges are above the mean density of background sources and reveal 
that these sources are concentrated about the centers of these galaxies. These trends are also seen in NGC~4374, but 
at lower significance. There is no clear evidence for an excess above the background in NGC~4636, except possibly for 
the reddest sources at intermediate radius. This is in agreement with the impression given from Figure~\ref{fig:images}, 
where the sources do appear concentrated towards the center of the galaxy, especially those within the intermediate color
range. Some fraction of these UV-bright sources must therefore be associated with
the target galaxies and cannot simply be an overlap in background sources.

\section{Source Identification}
\label{sec:ids}

A variety of sources emit in the UV, including old, horizontal branch and p-AGB 
stars, young stars, helium burning stars on the blue loop, open and globular clusters, AGNs, and
background galaxies. It is therefore crucial to determine the expected colors and magnitudes of 
each of these sources and address the likelihood of their detection in our targeted galaxies,
which we describe in the following subsections. 
As discussed in \S\ref{sec:cmds}, we have already eliminated globular clusters from our sample by restricting
our analysis to sources
with $\mathrm{F225W-F336W}<1$, as sources with F225W-F336W greater than $1$ are much too red to be 
the young stars and open clusters that we are searching for. Globular clusters are therefore 
omitted from the discussion below.

\subsection{Main Sequence Stars}
\label{sec:stars}

We used stellar evolutionary tracks from {\it BaSTI} \citep{2004Pietrinferni}, which were derived specifically 
for WFC3 filters, to construct a Hess diagram assuming a Salpeter IMF and 
constant SFR over the past 100~Myr. The tip of the main sequence is clearly visible in Figure~\ref{fig:cmds}, spanning
$-1.1 < \mathrm{F225W}-\mathrm{F336W} < -0.7$, 
and is where young, hot stars would be expected on the CMD if
they were present in our targets. In all but one of the targeted galaxies there is at least one detected source 
within this color range, making it entirely possible that they are young O or B stars. For reference, main sequence O stars range from $M_\mathrm{F225W}=-8.8$ to $-6.8$, while main sequence B stars rapidly become too faint to be detected (e.g., a B2V star has $M_\mathrm{F225W}=-4.8$).

It is clear in the CMDs, however, that the majority of sources are not located on the main sequence, but 
instead are mostly 
redder in color. To determine whether main sequence stars could be found at these colors, we estimated the 
effect of metallicity on the location of the main sequence. We created Hess diagrams using $\alpha$-enhanced 
models with $Z=0.0001$, $0.01$, $0.03$, and $0.04$. Higher metallicities indeed shift the evolutionary tracks 
to the red, but the magnitude of the effect is small: e.g., using $Z=0.04$ ($\approx 2.5 Z_\odot$) instead 
of solar metallicity results in a shift of the main sequence of +0.1 in color.

Another possibility that could account for the appearance of the sources being redder than that of the expected
location of the main sequence is the presence of dust. It is possible that there is dust at the center of some
or all of our targeted galaxies, and indeed is evident by obvious dust lanes that are visible in our data of NGC~4697 and NGC~4374. 
However, these visible dust lanes are limited to the very central regions of the galaxy (within a radius of $7\arcsec$ for
NGC~4374, the galaxy with the most prominent dust features), and most of our targets are much farther from
this region. So, although some sources may appear redder due to dust, it is unlikely that this is the case for
all of the targets between $-0.7 < \mathrm{F225W}-\mathrm{F336W} < 1.0$. 
We have also corrected for foreground reddening based on the dust maps of \citet{1998Schlegel}.

We conclude that if the sources lie on the grayscale of the Hess diagram in Figure~\ref{fig:cmds}, then  
by definition the sources can be individual stars with masses of $<30$~M$_\odot$, but that none or very few of the 
redder sources are individual stars.
It is possible, however, that they represent a population of open star clusters, a theory we investigate
in \S \ref{sec:clusters}.

\subsection{Star Clusters}
\label{sec:clusters}

If stars are forming in these nearby elliptical galaxies, it is likely that $70-90\%$ of the stars form
embedded in clusters instead of individually \citep{2003Lada}.
Although zero-age clusters are dominated by the most massive star, and therefore would lie on the grayscale in
Figure~\ref{fig:cmds}, they quickly evolve to redder colors and become fainter as they lose their bright O and
B stars.

\begin{figure*}
  \plottwo{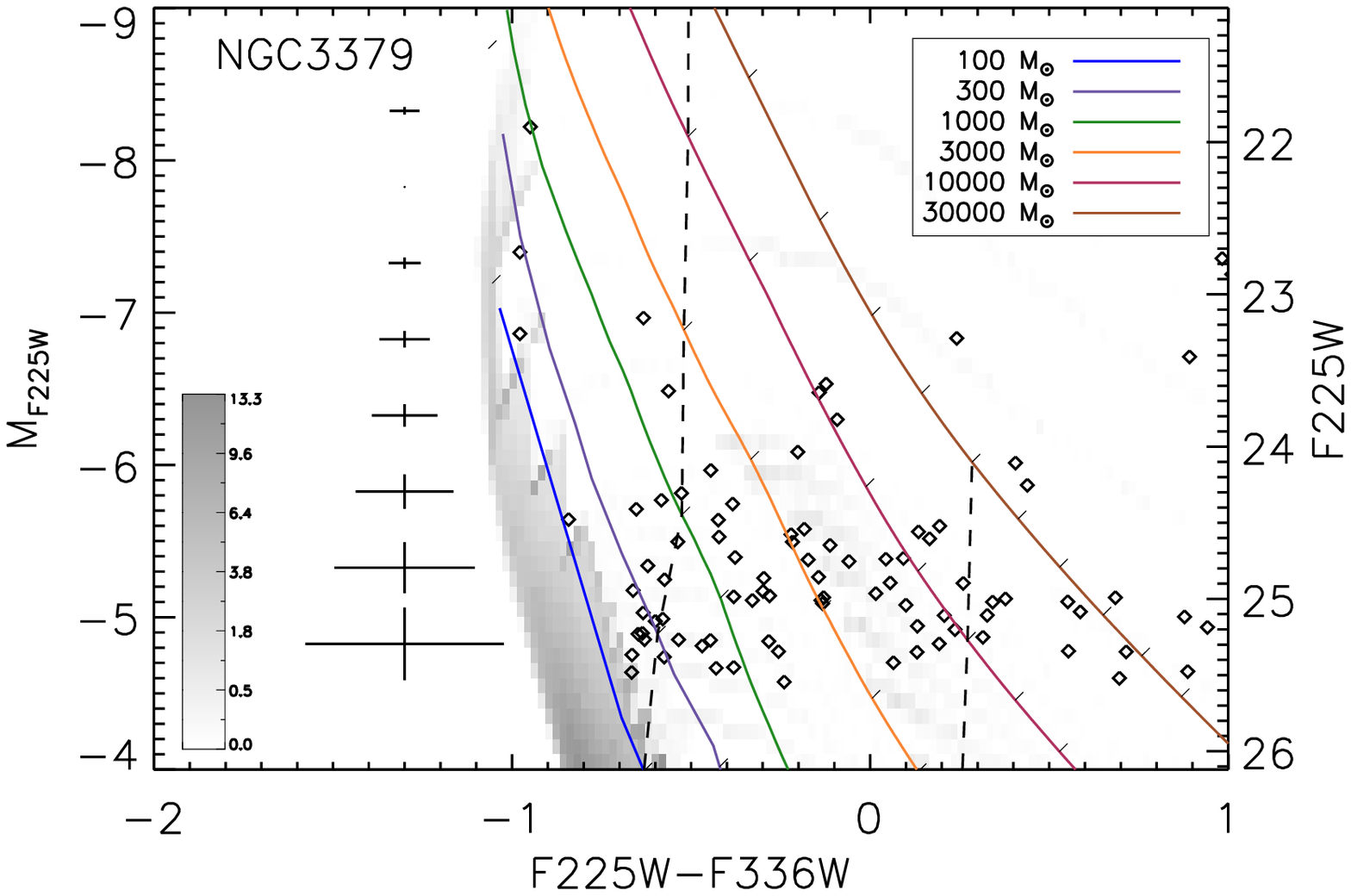}{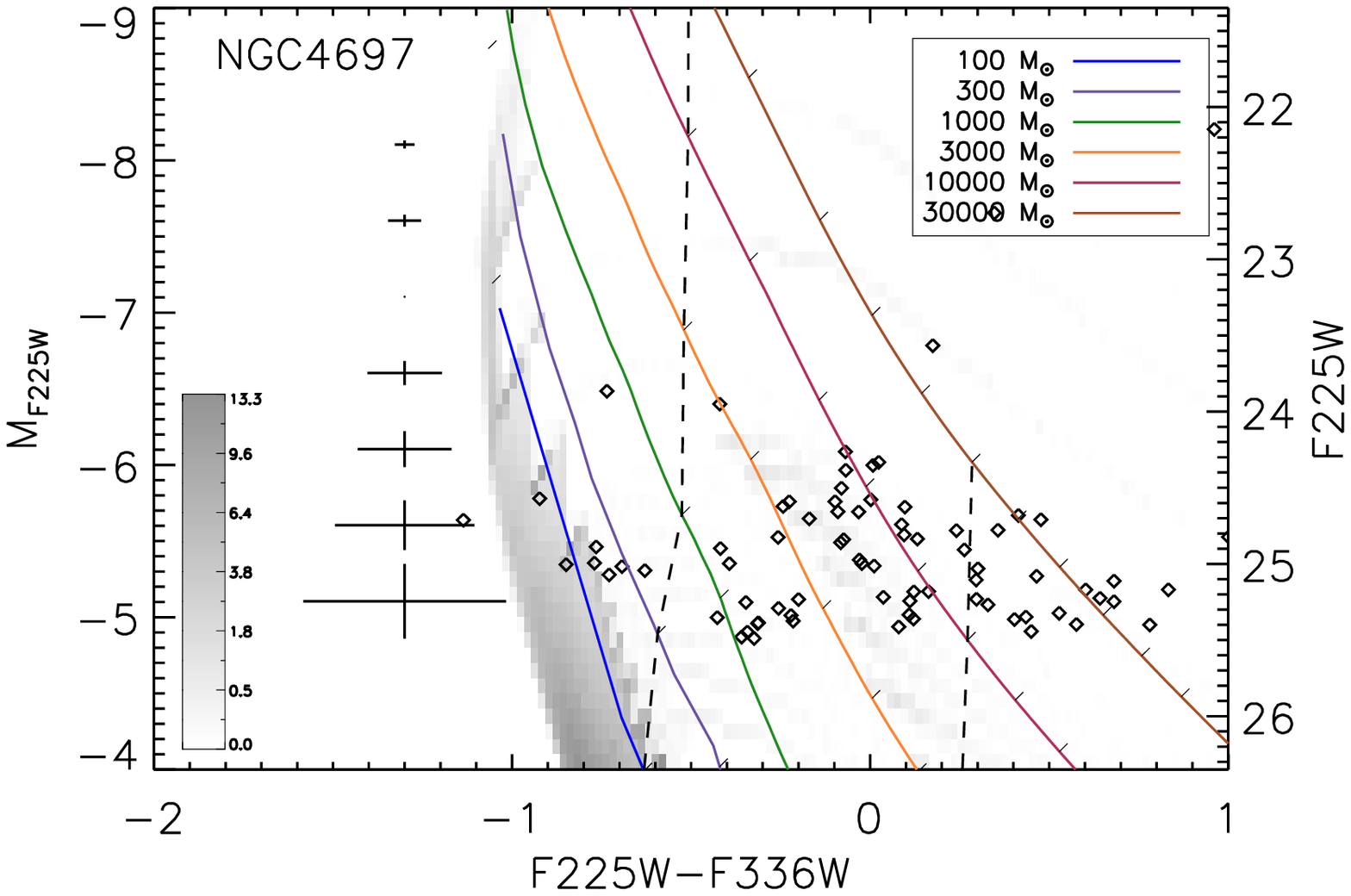}
  \plottwo{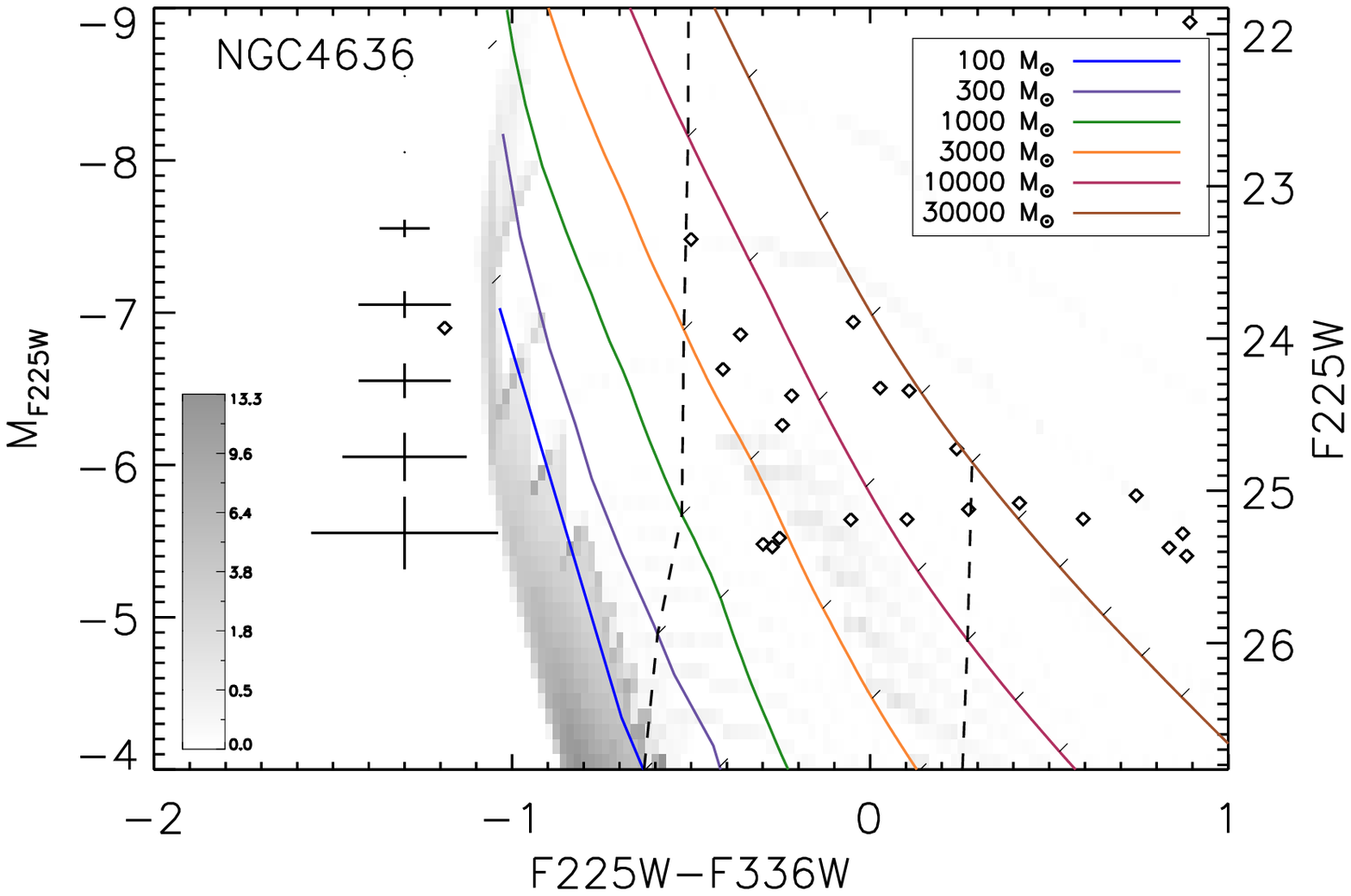}{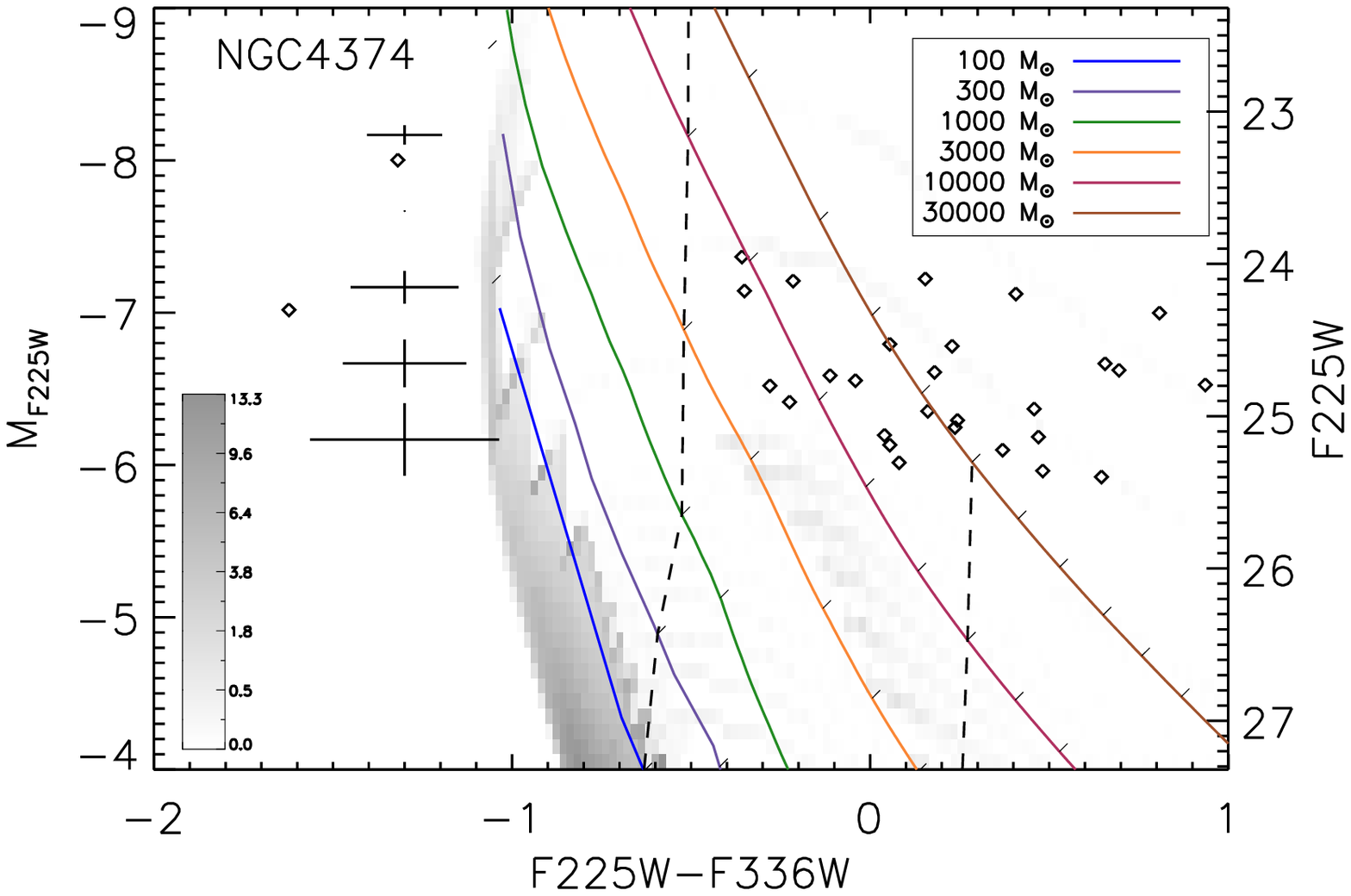}
    \caption{CMDs with evolutionary tracks of clusters overlaid. Cluster tracks assume a Salpeter IMF and are derived from the
      {\it BaSTI} stellar evolutionary tracks \citep{2004Pietrinferni}.  
Tick marks indicate ages of 0, 100, 200, 400, 600, 800, 1000, 1200, 1400, 1600, 1800, and 2000~Myr, while dashed lines indicate
100 Myr (left) and 1 Gyr (right).
Magnitude limits restrict the age at which we can detect lower mass clusters.
These CMDs have been corrected for background contamination.} 
    \label{fig:cmdsclustertracks}
  \end{figure*}

A CMD for each galaxy is shown in Figure~\ref{fig:cmdsclustertracks} with evolutionary tracks of open clusters
overlaid. The evolutionary tracks were determined assuming a Salpeter IMF and using {\it BaSTI} stellar 
evolutionary tracks \citep{2004Pietrinferni}, as a funtion of cluster mass, assuming that stars die instantly 
once they evolve off of the main sequence.
The main sequence lifetimes, $\tau_{\mathrm{MS}}$, are assumed to follow:
\begin{equation}
\tau_{\mathrm{MS}} = 10^{10}~\left(\frac{M}{M_{\sun}}\right)^{-2.5}~\mathrm{yr}
\end{equation}
which is a good approximation to the lifetimes in Figure 4 of \citet{1992Schaller} over the 
$30~\mathrm{Myr} \la \tau_{\mathrm{MS}} \la 2~\mathrm{Gyr}$ range of relevance.
Tracks for clusters with masses ranging from $10^2$~M$_\odot$ to $3\times 10^4$~M$_\odot$ are overlaid.
These CMDs have also 
been corrected for background contamination, where points on the CMD were matched to the distribution
of sources in the background CMD, which was scaled to the area of one WFC3 field, have been removed for a clearer
picture of the intrinsic CMD. As this procedure is statistical in nature, it should not be used when 
examining individual sources, but it gives an accurate picture of the effect of background contamination on the CMD.
Cluster tracks are the same for all four galaxies.

If the sources are open clusters, their distribution in the CMD reflects the cluster mass function, the rate 
at which clusters traverse the CMD, and the recent star formation history (SFH). For a constant SFH, the 
first two factors approximately cancel out at a given absolute magnitude: high mass clusters are produced 
less often than low mass clusters, but evolve more slowly (and therefore ``pile up'' on the CMD) at a given 
luminosity because they are later in their evolution. For example, if the cluster IMF slope is $-2$ 
\citep[e.g., see][]{2009Gieles}, clusters with masses between $10^2$~M$_\odot$ to $3\times 10^2$~M$_\odot$ are 10 
times more numerous as $10^3$~M$_\odot$ to $3\times 10^3$~M$_\odot$ clusters, but evolve 14 times faster between 
$-6 < \mathrm{F225W} < -5$, where most of the sources are located. We therefore expect star clusters to be 
approximately uniformly distributed in color at a given magnitude, as observed.

We estimated the disruption timescale of the clusters using equation (9) in \citet{2005Lamers}, which scales 
as cluster mass to the 0.62 power and as the inverse square root of the ambient density. This latter quantity, 
which is dominated by the baryonic component at these small radii, we estimate using the \citet{1983Jaffe} 
model, which provides a good model for elliptical galaxy light profiles. For a mass of $2\times 10^{10}$~M$_\odot$ 
and an effective radius of 2 kpc, typical of the target galaxies, the ambient density at the effective radius is 
$\sim 0.05$~M$_\odot$~pc$^{-3}$, and the cluster disruption timescale is $>1$~Gyr for $10^4$~M$_\odot$ clusters, 
and $\ga 100$~Myr for even $10^2$~M$_\odot$ clusters. We would therefore expect that clusters from present-day 
star formation have not yet disrupted. Moreover, disruption is not instantaneous, and clusters in the process 
of dissolving may well still appear as point sources at these distances. 

To determine the radial extent of the sources, it was necessary to stack all sources to increase the 
signal-to-noise. A variety of stacks were created spanning different color ranges to ensure that there was 
no dependence of the radius on color; color had no effect on the determined radii. Stacks were scaled by 
the inverse of the flux, making all sources weighted equally regardless of magnitude. We used Tiny Tim 
HST PSF modeling software \citep{2011Krist} to determine the PSF for WFC3 F225W filter, resulting in a PSF 
with $FWHM=1.7$. The stacked radius was more extended than the PSF, with a $FWHM=2.5$. The radius of the 
stacked sources for the F336W filter was also more extended than the PSF derived using Tiny Tim. While 
the minimum resolvable radius is 3.4~pc, we determine that the mean cluster radius is 5.0~pc (based on 
the radius of the F225W cluster stack). This is consistent with the sources being open clusters; for example, 
the median cluster radius in the Catalogue of Open Cluster Data \citep[COCD;][]{2005Kharchenko} is 4.3~pc.

\subsection{Post-Main Sequence Stars}
\label{subsec:pmss}

\begin{figure}
  \plotone{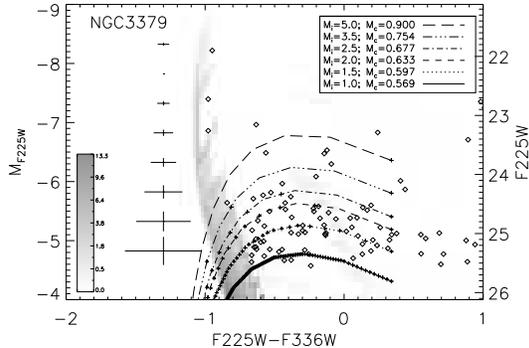}
  \caption{CMD of UV-bright sources in NGC~3379, our closest target and therefore deepest CMD.
    Overlaid curves represent p-AGB evolutionary tracks from \citet{1994Vassiliadis}. Tracks are labelled according 
    to the stellar model initial mass, $M_i$, and the final core mass, $M_c$. Ticks are plotted every 100 yr.
    A main sequence turnoff mass of $M_i=1.0$ corresponds to a 10~Gyr population, $M_i=1.5$ corresponds to an age 
    of $<4$~Gyr and brighter tracks are produced by turnoff stars from even younger populations. 
    The bottom track is therefore the only place we would expect to see a
    population of p-AGB stars from the dominant old population, and our
    sources are too bright to be these p-AGB stars. This
    CMD has been corrected for background contamination.}
  \label{fig:cmdspagbtracks}
\end{figure}

Post-main sequence stars, such as hot horizontal branch and p-AGB stars, contribute much of the UV light from 
galaxies \citep{1999OConnell}. 
Typical p-AGB stars detected in M32 and modeled from stellar evolutionary tracks by \citet{2008Brown} are too
faint by several magnitudes relative to our data to contaminate our sample. Theoretical considerations also make it 
unlikely that the sources we detect are old p-AGB stars. The p-AGB evolutionary tracks of \citet{1994Vassiliadis} 
have been overlaid on our deepest CMD in Figure~\ref{fig:cmdspagbtracks}, where the bolometric correction has been 
computed assuming a blackbody spectrum. Although the colors of our sources are consistent with being p-AGB stars, the 
magnitudes are not: only stars with initial masses $\ga 1.5$~M$_\odot$ could be bright enough to be the sources we 
detect, which implies that they would have to belong to a reasonably young ($<4$~Gyr) population. Moreover, such 
massive stars live very short lifetimes as p-AGB stars, as seen by the length between 100 yr tick marks on 
Figure~\ref{fig:cmdspagbtracks}. 
As these stars spend more time blueward of the main sequence than redward of it, the complete absence of sources blueward of the main sequence is inexplicable if they are p-AGB stars, but is a natural consequence if they are star clusters that evolve redward from locations slightly to the red of the main sequence. Horizontal branch stars are similar in
temperature to p-AGB stars, but significantly fainter, and so cannot be the sources we detect.

Helium burning stars on the blue loop are expected to have colors redder than the main sequence, though that
number should be low based on the grayscale probability distribution derived from stellar evolutionary tracks, 
i.e., the blue loops are the regions seen in Figure~\ref{fig:cmds} between 
$-0.5 \lesssim  \mathrm{F225W-F336W} \lesssim 0.2$ at the magnitudes we are detecting.
The grayscale probability distribution clearly
illustrates that few stars are expected in this region, i.e., only the helium burners on the blue loop would
be in these color ranges, and even then they are so short-lived that they would be very unlikely. Also, if these
were blue loop stars, many more stars on the main sequence would be expected based on the Hess diagram, i.e., at 
magnitudes brighter than $M_\mathrm{F225W}=-4.6$, we would expect to see 19 times more stars on the main sequence 
than on the blue loop, though we only detect a few sources on the main sequence while detecting many sources
redward of this.

\subsection{Point-like Background Galaxies and AGNs}

Contamination from point-like background galaxies and AGNs is unavoidable. However, the amount of contamination
can easily be estimated by using a blank control field to get an estimate of the colors, magnitudes, and number 
of sources expected in such a field. In section \ref{bcf} we described eight blank fields observed with
WFC3 using the same filters as our observations, which we photometered to derive a background sample.
After scaling the total area of the combined blank fields to the area of one WFC3 field, we determined
how many background sources would be expected on our CMDs and in what position on the CMD they would be in.
We then matched these background sources to the most likely counterpart in our CMDs based on their magnitudes 
and colors, and then subtracted them to determine what the CMD corrected for background contamination would 
be. The resulting CMDs of this process can be see in Figure~\ref{fig:cmdsclustertracks}, which demonstrates that
although a background sample has been removed from the CMD, there are still numerous sources within the CMD
that are unaccounted for, in each of our galaxies. Although galaxy clustering can change the magnitude of the 
background counts, the radial dependence of the source distribution clearly demonstrates that the majority 
of detected sources are contained in the target galaxy and are therefore not point-like background galaxies or 
AGNs. An estimate of the variance in the background can be obtained by looking at the field-to-field 
dispersion in the eight blank fields; the mean number of sources is 11.6, with a standard deviation of 5.8. In 
contrast, the number of sources detected within the color range of our study in NGC~3379, NGC~4636, NGC~4697, 
and NGC~4374 are 98, 31, 87, and 41, respectively.

\subsection{Surface Brightness Fluctuations}
\label{sec:sbf}

It is important to address the likelihood of false detections due to surface brightness fluctuations (SBFs),
which will not only result in an increased number of detections, but can also mimic a radially
concentrated population. Models by \citet{1993Worthey} determine the characteristic SBF magnitude in
F284W and the optical U band for a variety of old stellar populations. In the case where there is both an extended/blue
horizontal branch and a low-mass p-AGB star population contributing to the SBFs, the characteristic SBF
magnitude will be the luminosity-weighted mean luminosity of the populations. Using their low-mass p-AGB
case as a worst case scenario, $(\overline{\mathrm{F284W}}, \overline{\mathrm{U}}) = (1.2, 2.7)$, which, when converted
to the filters we are interested in, implies $(\overline{\mathrm{F225W}}, \overline{\mathrm{F336W}}) = (-0.3, 2.7)$.
As the deepest limit in our data is -4.5, this corresponds to nearly a $50\sigma$ fluctuation, which is extremely
unlikely, particularly over the WFC3 field of view. We therefore rule out the possibility that these sources
are surface brightness fluctuations.

\section{Results}
\label{sec:sfr}

As the detected sources are most likely individual main sequence stars and star clusters, we can deduce their
individual ages and masses from evolutionary tracks. The properties of our sample of UV-bright sources 
for each galaxy are presented in Table~\ref{tab:basic}, where the galaxy ID is in column 1, right 
ascension ($\alpha$) and declination ($\delta$) are in columns 2 and 3, apparent F225W and F336W 
magnitudes are in columns 4 and 5, and color (F225W$-$F336W) is in column 6. The derived cluster ages 
and masses are also presented in Table~\ref{tab:basic} in columns 7 and 8. Errors for the cluster 
properties were determined by Monte Carlo sampling the error ellipse in the CMD around each cluster, 
and represent the 16th and 84th percentiles of the Monte Carlo distribution ($1 \sigma$).
Because the clusters evolve faster at younger ages, the errors are asymmetric and there is expected 
to be a net bias in the age and mass determinations. We estimate the effect of this by examining the 
mean of the Monte Carlo error distribution. We find that the mean age after accounting for the 
photometric error is $\sim 30$~Myr older than the input age for clusters younger than $1$~Gyr (the 
method is not well defined for older clusters because the Monte Carlo points run off of the calculated 
evolutionary tracks), and the recovered mean mass is larger than the input mass by a factor falling 
from ~3 at $100$~M$_\odot$ to $\sim 1.5$ at $1000$~M$_\odot$ and is negligible at $10^4$~M$_\odot$. This 
systematic error should be kept in mind when interpreting the cluster properties.

\begin{deluxetable*}{lrrccrcc}
\tabletypesize{\scriptsize}
\tablecaption{Properties of UV-Bright Clusters \label{tab:basic}}
\tablehead{
\colhead{Galaxy}&\colhead{$\alpha$}&\colhead{$\delta$}&\colhead{F225W}&\colhead{F336W}&\colhead{F225W$-$F336W}&\colhead{Mass}&\colhead{Age}\\
\colhead{}&\colhead{}&\colhead{}&\colhead{}&\colhead{}&\colhead{}&\colhead{($M_{\odot}$)} & \colhead{(yr)}}
\startdata
NGC~3379 & 10:47:43.078 & +12:35:20.59 & $25.04 \pm 0.19$ & $24.94 \pm 0.08$ & $0.10 \pm 0.21$ & $7.04^{+6.56}_{-3.78} \times 10^{3}$ & $7.06^{+3.24}_{-2.66} \times 10^{8}$\\
  & 10:47:44.170 & +12:34:42.50 & $25.37 \pm 0.25$ & $26.03 \pm 0.19$ & $-0.67 \pm 0.32$ & \nodata & \nodata\\
 & 10:47:44.325 & +12:34:34.47 & $25.03 \pm 0.19$ & $25.16 \pm 0.09$ & $-0.13 \pm 0.21$ & $3.07^{+3.67}_{-1.76} \times 10^{3}$ & $4.00^{+2.87}_{-1.49} \times 10^{8}$\\
 & 10:47:44.539 & +12:34:48.99 & $24.98 \pm 0.18$ & $25.37 \pm 0.11$ & $-0.38 \pm 0.21$ & $1.17^{+1.69}_{-0.79} \times 10^{3}$ & $2.19^{+2.52}_{-0.99} \times 10^{8}$\\
 & 10:47:44.621 & +12:34:21.47 & $25.48 \pm 0.30$ & $26.15 \pm 0.21$ & $-0.67 \pm 0.37$ & \nodata & \nodata\\
 & 10:47:45.475 & +12:35:01.37 & $24.60 \pm 0.14$ & $24.44 \pm 0.05$ & $0.17 \pm 0.15$ & $1.30^{+0.70}_{-0.51} \times 10^{4}$ & $8.23^{+2.02}_{-1.98} \times 10^{8}$
\enddata
\tablecomments{
Table \ref{tab:basic} is published in its entirety in the electronic edition of the {\it Astrophysical Journal}.
A portion is shown here for guidance regarding its form and content. 
Sources that lie outside the cluster evolution tracks do not have mass or age listed. Sources where the upper age error lies outside the cluster evolution tracks are listed as lower limits.
}
\end{deluxetable*}

\begin{figure*}
  \plottwo{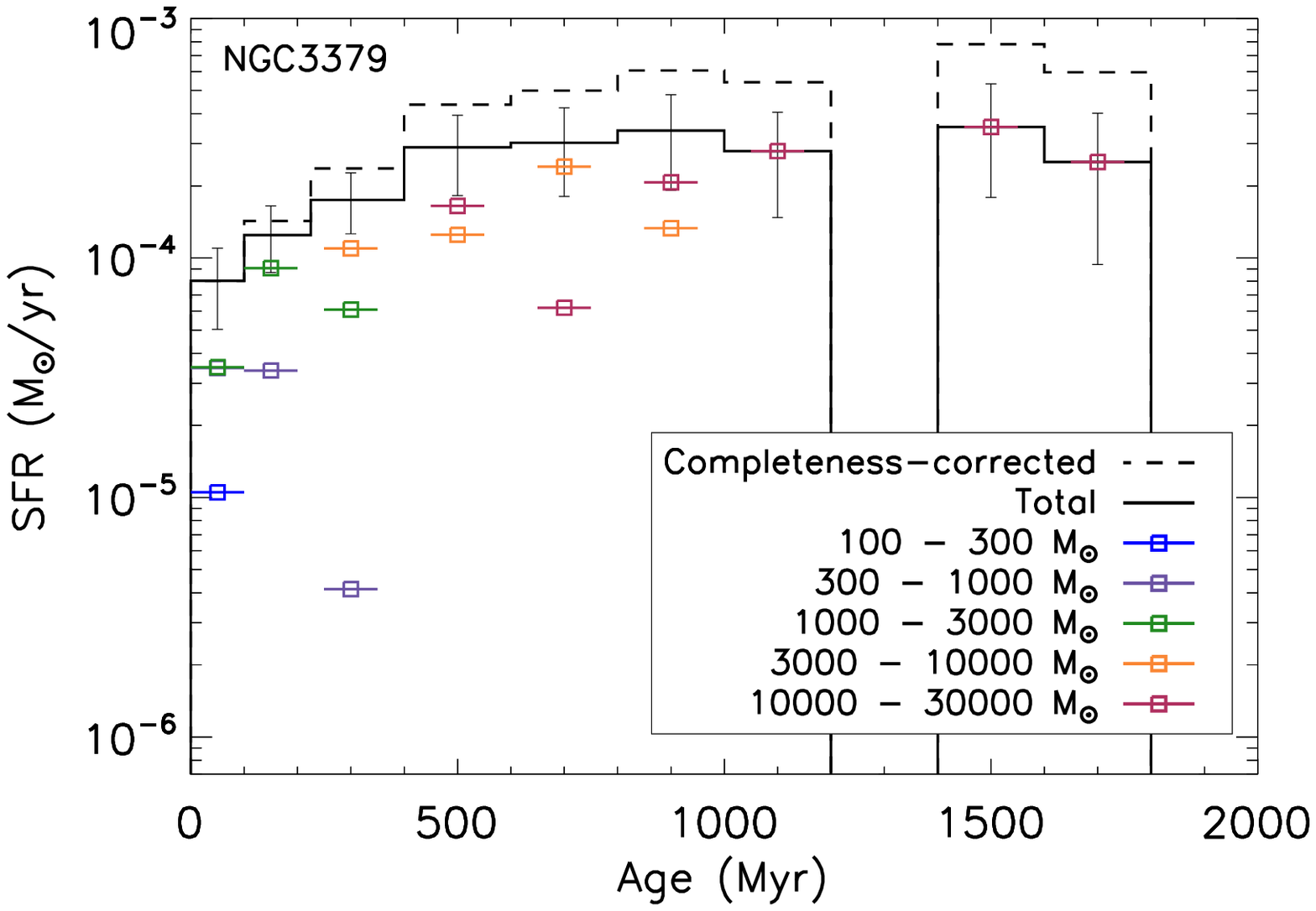}{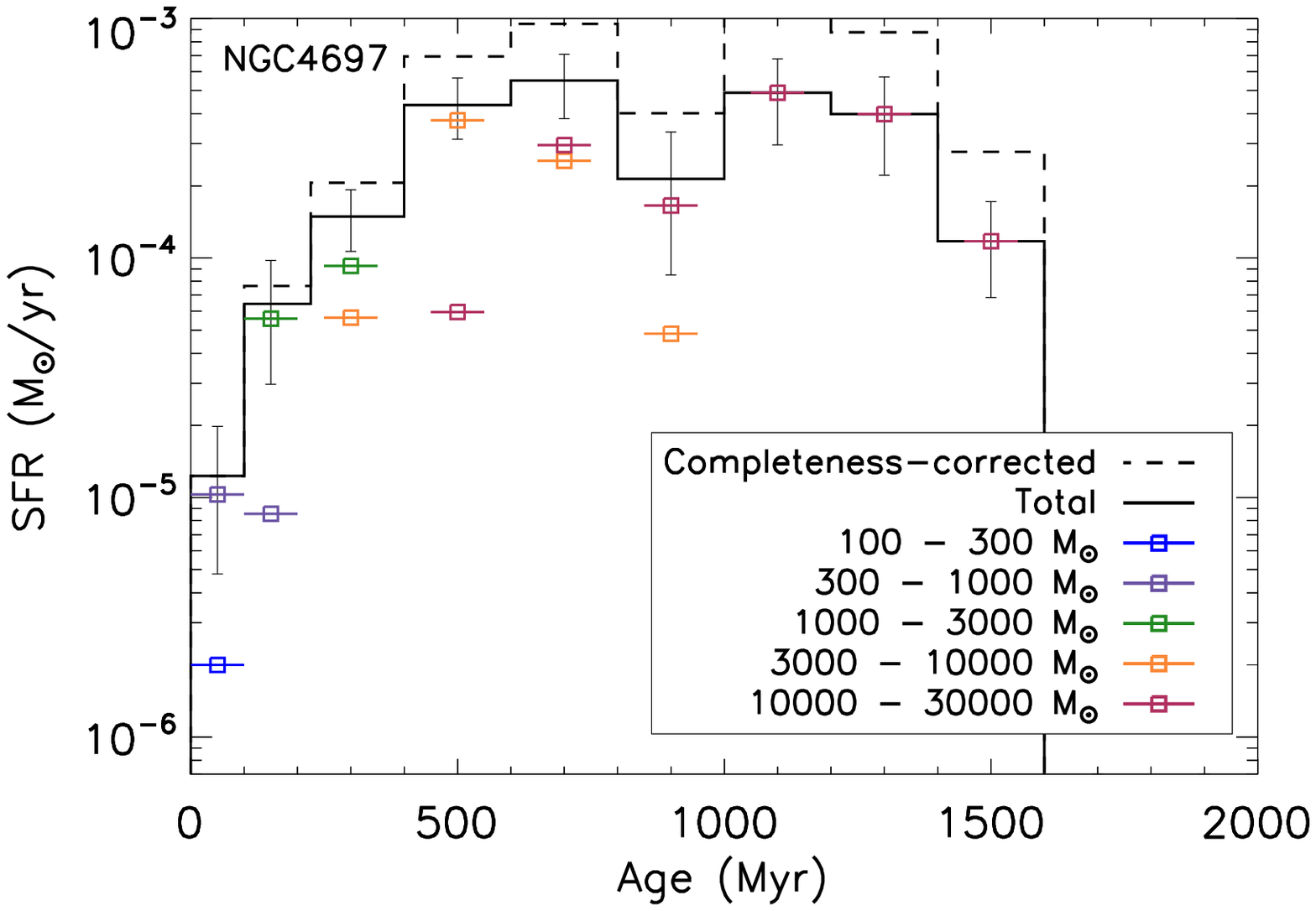}
  \plottwo{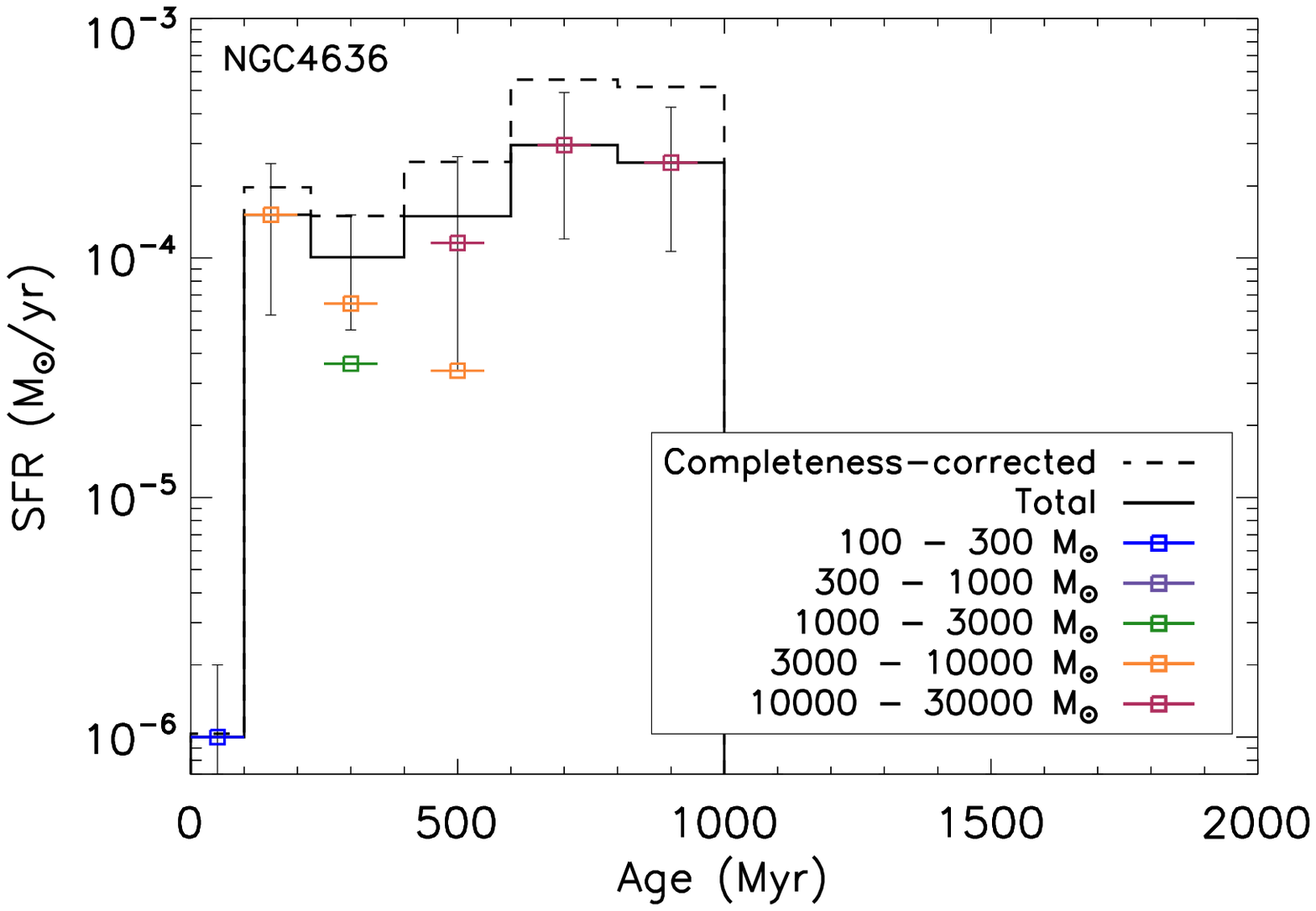}{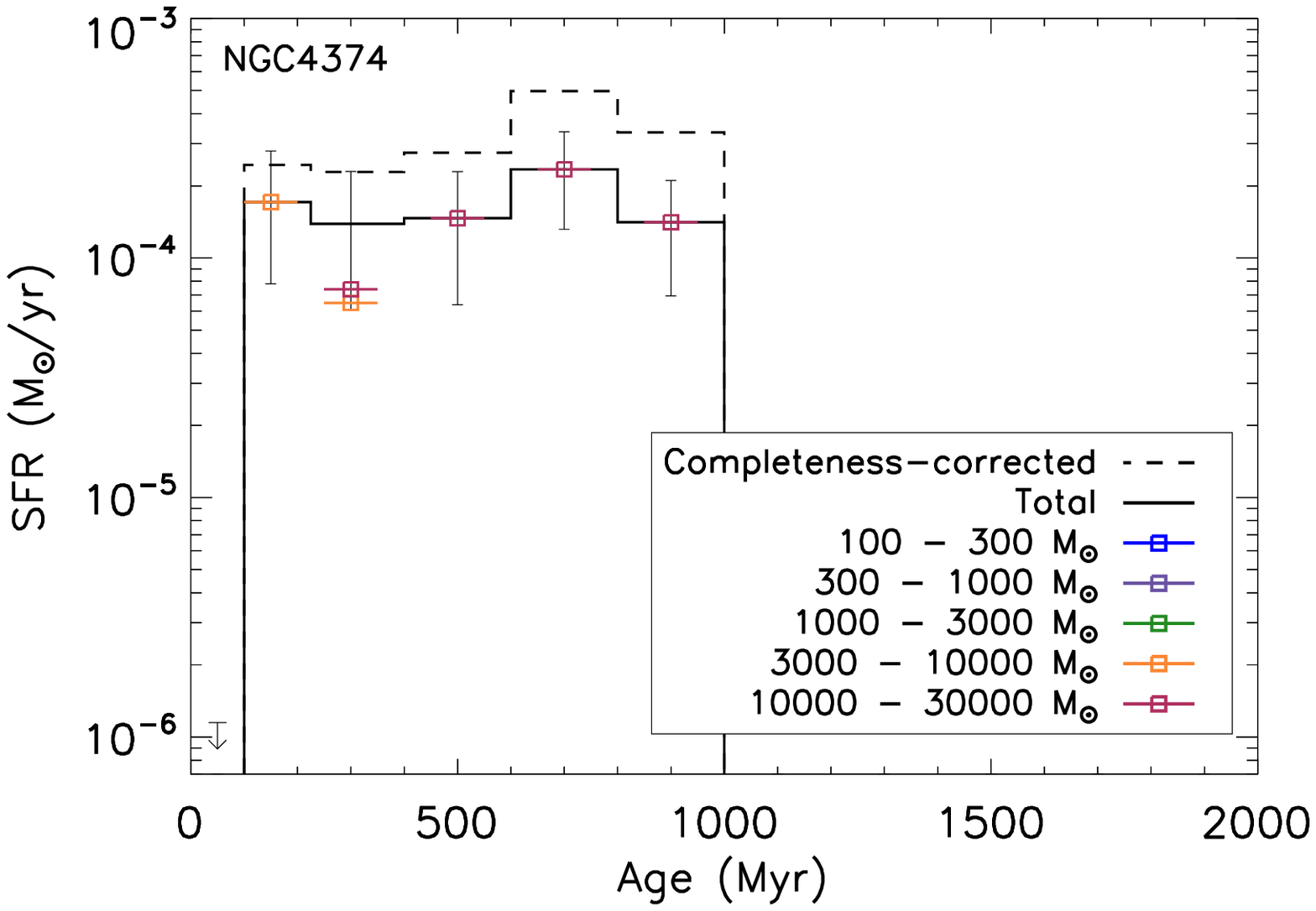}
  \caption{Star formation rate as a function of cluster age. Individual points represent
    the star formation rate contributed by clusters as a function of mass, while the solid line
    represents the total star formation rate per bin. There is ongoing star formation in all four
    of the observed galaxies, though more in some than in others. These have
    been corrected for background contamination.  The completeness
correction assumes that the cluster catalog is complete to the mass where the cluster age tracks
cross the typical magnitude limit for each galaxy, using a cluster initial mass function that is a
power law of slope $-2$ from $10^2$ to $10^6$~M$_\odot$.
Error bars are plotted for the ``Total'' line, and are based on bootstrap resampling of a Poisson-distributed
number of clusters within each age bin.
The relative errors on the completeness-corrected SFRs are identical.}
  \label{fig:sfrvsage}
\end{figure*}

Assuming that all star formation results in an extant cluster, we can add together the mass of clusters
within a given age range to estimate the star formation rate at that time.
Star formation rate as a function of cluster age is shown in Figure \ref{fig:sfrvsage}, where the
individual points represent the star formation rate contributed by clusters as a function of mass and the solid line
represents the total star formation rate per bin. The cluster age errors are typically $\lesssim 100$~Myr for 
ages $<200$~Myr, rising to $200-400$~Myr for older clusters. We have therefore adjusted the bin sizes in 
Figure~\ref{fig:sfrvsage} to be similar to the typical errors.
There is ongoing star formation in all four of the observed 
galaxies, and there is variation between galaxies. As the data get shallower, there is a bias that lower mass 
clusters are lost first. We have applied a completeness correction, which assumes that the cluster catalog is 
complete to the mass where the cluster age tracks cross the typical magnitude limit for each galaxy, using a 
cluster initial mass function that is a power law of slope $-2$ from $10^2$ to $10^6$~M$_\odot$
\footnote{The total mass has a very mild logarithmic dependence on the upper and lower limits
of the mass function, so the choice of these particular values does not introduce a large
uncertainty.}. As can be seen 
in Figure~\ref{fig:sfrvsage}, applying the completeness correction does not have a dramatic effect on the implied 
star formation rates, and we are therefore directly constraining the star formation rate with the observed clusters.

The derived SFRs take background contamination into account by subtracting those sources that best match 
the distribution of the scaled control background field. 
At ages $> 10^9$~yr, the colors of star clusters become more sensitive to cluster metallicity
than to cluster age --- changes of $0.1$~dex in metallicity become enough to move the clusters
into different age bins --- and therefore the ages of clusters we list as older than this
should be considered with care.
Similarly, the details of the star formation histories
beyond 1~Gyr are very uncertain.
As noted above, photometric errors cause a mean overestimate in cluster masses at the low mass end, resulting
in a tendency to overestimate the star formation rate contribution by those clusters. However, 
only for the lowest-age bin in NGC~4636 do the lowest-mass clusters contribute a significant amount to 
the total estimated star formation, so this should have a minimal effect on the derived star 
formation rates. The mean age overestimate due to photometric errors is significantly smaller than 
the size of the age bins, so it should not have a significant effect.

In three of the four galaxies, we detect current star formation at rates that range from
$1.0\times 10^{-6}$~M$_\odot$~yr$^{-1}$ for NGC~4636 to $8.0\times 10^{-5}$~M$_\odot$~yr$^{-1}$ for
NGC~3379 (note that because the completeness is different between galaxies, we quote completeness-corrected
values here, although the difference is minimal for the present-day SFR); in NGC~4374, the lack of 
detected $<100$~Myr clusters corresponds to an upper limit of $1.15\times 10^{-6}$~M$_\odot$~yr$^{-1}$.
Star formation within the past Gyr is detected in all four galaxies, 
with average rates somewhat larger, ranging from $2.9\times 10^{-4}$~M$_\odot$~yr$^{-1}$ for NGC~4374 up 
to $4.6\times 10^{-4}$~M$_\odot$~yr$^{-1}$ for NGC~4697.
Encouragingly, these numbers are of the same order as the average star formation rates found by
\citet{2010DonovanMeyer}
for early-type galaxies that had no other evidence for star formation, based on total GALEX NUV emission. It
therefore seems likely that the emission they see is the unresolved sum of the sources we detect with HST.

\begin{figure}
  \plotone{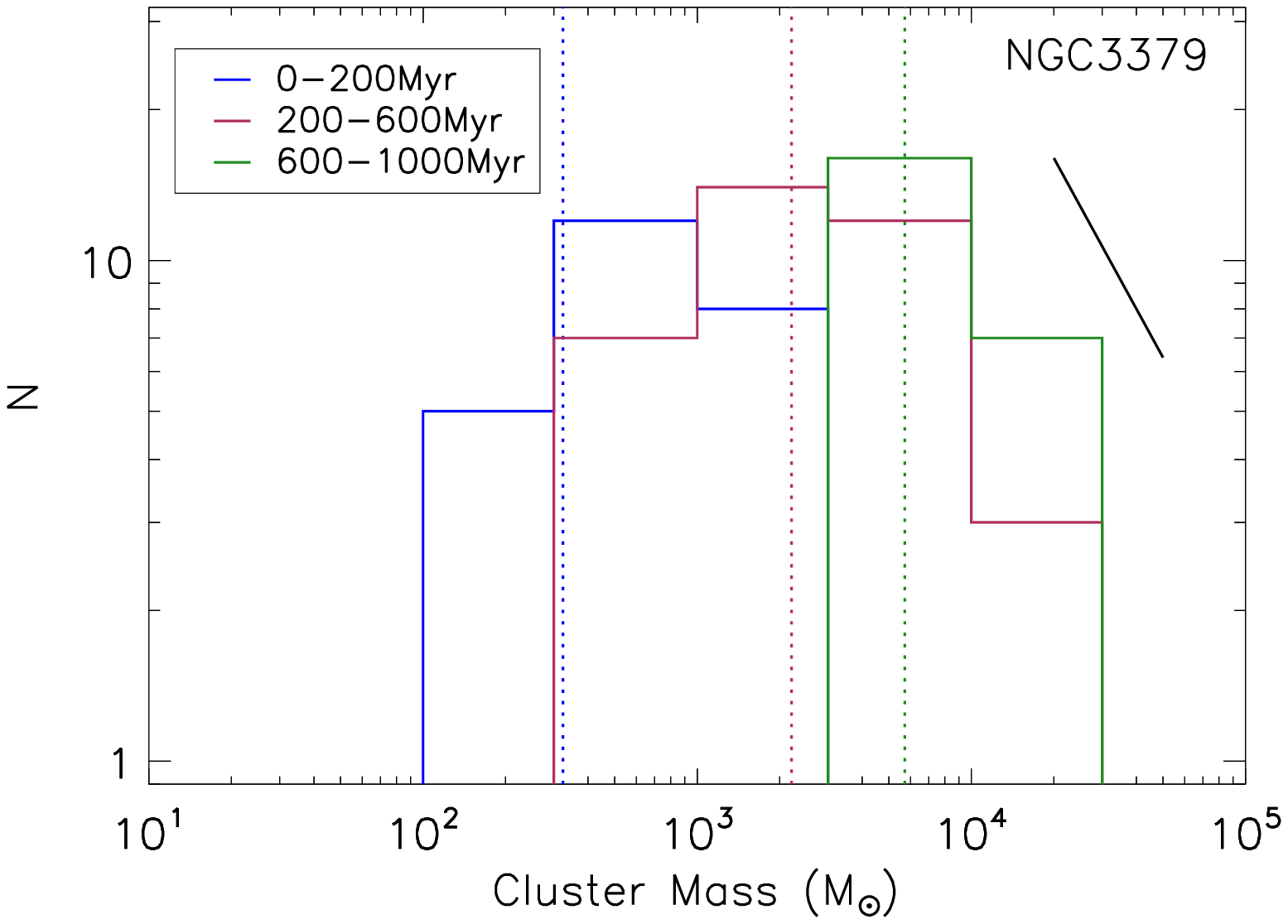}
  \caption{Cluster mass function for NGC~3379 for three different age ranges, with the completeness limit shown 
    in each case as the vertical dotted line. The black diagonal line shows a power law of slope $-2$ for reference.
}
  \label{fig:clustermassfunction}
\end{figure}

The cluster mass function presents another projection of the cluster data and is shown in 
Figure~\ref{fig:clustermassfunction} for the nearest galaxy in our sample, NGC~3379. The 
data for the remaining galaxies are too shallow to draw strong conclusions on the mass 
function, but they are consistent with what is seen in NGC~3379.
Because the mass function is complete to different masses at different ages, we have plotted 
the mass function for clusters in three different age ranges, with the completeness limit shown in each case as 
the vertical dotted line. The black diagonal line shows a power law of slope $-2$ for reference. 

One notable feature in Figures~\ref{fig:sfrvsage} and \ref{fig:clustermassfunction} is the lack of very massive
($> 3000$~M$_\odot$), very young ($< 300$~Myr) clusters. Such clusters would not be too 
faint to be detected, and they should not have been disrupted. We can imagine several 
potential explanations for this, however most are not viable upon further examination.
For instance, particularly luminous sources could have diffraction spikes that are
artificially broken up by the source finding algorithm; however, the successful 
detection of more luminous sources at redder colors refutes this explanation. 
Moreover, visual inspection of those few detections that dolphot removed for being 
extended revealed all to be truly extended objects rather than diffraction spikes.
The systematic age overestimate from photometric errors also cannot be the explanation, 
as this would be more important for low mass clusters than high mass clusters.
Another potential explanation is that there are no high mass clusters at any age and the 
apparent old high mass clusters are actually blue loop stars. However, as argued in 
Section~\ref{subsec:pmss}, the lack of observed main sequence stars precludes a large population of
blue loop contaminants. Furthermore, such a population would produce a peak in
Figure~\ref{fig:clustermassfunction} in the $3000-10000$~M$_\odot$, $200-600$~Myr bin, where no such
feature is seen. While it is true that the cluster evolution tracks are based on stellar
models that are poorly constrained in the UV, and could potentially be
systematically too blue resulting in systematically high age estimates,
this also cannot be the explanation because it would require the highest
mass stars to be significantly bluer than the lower mass stars.

The most likely explanation is that there is a physical reason why galaxies
in this sample have star formation that is not vigorous enough to
generate clusters more massive than $3000$~M$_\odot$. Given that the galaxies
are all very red ellipticals, this is entirely possible, particularly
as they have current star formation rates that are extremely low
even relative to their average over the past Gyr. This is augmented
by small number statistics -- at these low star formation rates, only
one very massive very young cluster per galaxy would be expected.

\begin{figure}
  \plotone{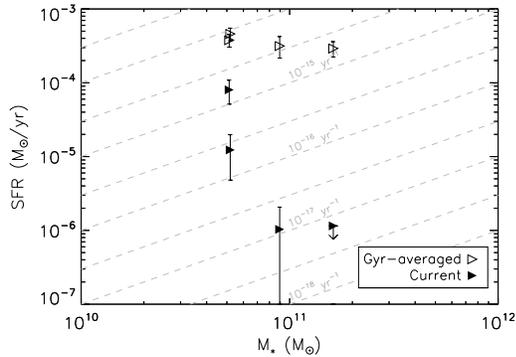}
  \caption{Star formation rate as a function of stellar mass for all four galaxies in our sample. Filled 
    symbols indicate the SFR over the past $100$~Myr, while open symbols are the average SFR over the past 
    Gyr. Gray lines of constant specific star formation rate are overlaid. The Gyr-aged SFRs are more
    or less independent of stellar mass, while the current star formation appears to decline with mass.
}
  \label{fig:sfrvsmass}
\end{figure}

Figure~\ref{fig:sfrvsmass} shows how the current and recent star formation of the galaxies vary as a function of 
their stellar mass, where the stellar masses were calculated from 2MASS LGA \citep{2003Jarrett} $K_{tot}$ magnitudes and the $M/L$ 
calibration of \citet{2001Bell} using the $V-K$ color from the RC3 $V_T$ magnitudes. The Gyr-averaged 
SFRs are more 
or less independent of stellar mass, while the current star formation appears to decline with mass. 
Lines of constant specific star formation rate (SSFR) are overplotted. The SSFR values range from 
$<8\times10^{-18}$~yr$^{-1}$ to $2\times10^{-15}$~yr$^{-1}$ at the present day, and $2\times10^{-15}$~yr$^{-1}$ to 
$10^{-14}$~yr$^{-1}$ averaged over the past Gyr.

\section{Discussion}
\label{sec:discussion}

This work presents the first derivation of star formation rates of true ``red and dead'' ellipticals via 
direct detection of point sources in the UV using the WFC3 on HST. Our targets have distances ranging from
10.6 Mpc to the Virgo Cluster at 18.4 Mpc, and three were chosen because they demonstrated physical 
evidence of possibly harboring young stars, in the form of PAH emission, OVI emission, and AGN activity, 
while the fourth was selected as the control galaxy with no star formation indicator. The detected 
sources are best explained as being individual star clusters with masses between $10^2$ and $10^4$~M$_\odot$ 
and ages less than $1$~Gyr: they are radially concentrated about the center of each galaxy, ruling out 
background contamination; they are too bright to be p-AGB stars from the dominant old population and 
too red as a population to be p-AGB stars from an intermediate-age population; and the absence of a large 
number of bright main sequence turnoff stars precludes a large enough population of blue loop stars to 
account for any but a tiny fraction of the observed sources. Our detections of young stars and star 
clusters in all four of our targeted ellipticals confirms that low-level star formation is
ongoing in these ``red and dead'' galaxies.

All of the observed galaxies have line index ages \citep{2010Kuntschner,2006Sanchez,2000Trager} that 
are uniformly very old. To estimate whether the very small amount of star formation we detect is consistent 
with these ages, we compared the total F336W light coming from our detected sources to the total U-band flux 
from RC3. For NGC~3379, where we go deepest down the young cluster mass function, only $0.03\%$ of 
the U-band light within the effective radius is coming from our sources. It is therefore consistent with the 
old ages determined spectroscopically. A current SFR that is $1$--$2$ orders of magnitude larger, which would 
still be immeasurable by other techniques but is easily ruled out for the galaxies in this sample, would be 
required to provide enough frosting to begin to contribute enough light at the age-sensitive wavelengths 
to alter the derived ages. Our data are also consistent with the models of \citet{2010Rogers}, which were 
performed to study the star formation history of NGC~4697 (amongst other ellipticals). Their results show that 
a simple stellar population model has an equally good fit to the data as their best fit frosting model, meaning 
that up to $3\%$ of the stellar mass in a younger component can be added without changing the fit. Our 
derived value is consistent with this result, as the young population we detect makes up a much smaller percentage 
than this.

The typical star formation rates of these galaxies are $\sim 10^{-5}$~M$_\odot$~yr$^{-1}$. 
The two closest galaxies, NGC~3379 and NGC~4697, where we have the best measurements of the SFR, both show current SFR at a
similar level, despite the presence of PAH emission from one and not the other. The errors are much
larger for the more distant galaxies with OVI emission, NGC~4636 and NGC~4374, but they also show consistent results. This perhaps
indicates that such emission is a poor tracer of the small-scale minute amounts of cold gas that must
be present to form stars. 

Comparison to the current stellar mass of the galaxies gives specific star formation rates of 
$\sim 10^{-16}~\mathrm{yr}^{-1}$ at the present day, which implies that stars younger than $100$~Myr provide
a frosting that consists of $10^{-8}$ of the total stellar mass. The Gyr-averaged specific star formation rate 
is somewhat higher, $\sim 10^{-14}~\mathrm{yr}^{-1}$, corresponding to $10^{-5}$ of the stellar mass.

There is no obvious structure to the spatial distribution of clusters beyond their general concentration 
towards the center of each galaxy, or any obvious correlation between the locations of the clusters and 
any other feature in the galaxy.

The star formation histories of the galaxies reveal a drop in the star formation within the past 300 Myr, 
particularly in NGC~3379 and NGC~4697 where the data are the deepest and the sources are most obviously 
physically associated with the galaxy. There are several possible origins for this, such as a contaminant 
that we have not properly accounted for that pollutes the region of the diagram where older clusters lie,
though this explanation is highly unlikely. 
Another possibility is that the cluster evolution tracks evolve too slowly in this region of 
the CMD. More sophisticated population synthesis models in the WFC3 UVIS bands would be required to test 
this hypothesis, which are not presently available. 
Or perhaps our selection of galaxies, which was not random but specifically targeted old red galaxies with no known 
star formation, picked out galaxies that are unusual quiescent at the present day. The youngest clusters are the 
easiest to see, easier than old clusters, 
but they are completely absent in 2 of the 4 galaxies. One of these galaxies, NGC~4374, hosts an AGN,
which may have recently heated the gas and made it unavailable for star formation. This hypothesis can be
tested with a larger and more representative sample of elliptical galaxies, which we hope to obtain.

These results provide a key observable for the amount of residual star formation in quenched early-type galaxies. 
By comparing the rate at which stellar mass transitions across the ``green valley'' to the stellar mass function 
of the red and blue galaxy populations, \citet{2007Martin} inferred a quenching timescale on the order of hundreds 
of Myr. However, because these small levels of star formation have minimal impact on the global galaxy color, the 
residual star formation had been unconstrained. These results will therefore provide useful input for models that 
attempt to measure the ages of the stellar populations of galaxies based on
their spectral features. In particular, it would be interesting to see whether frosting models such as those of 
\citet{2000Trager}, when using these SFHs, are able to reconcile the apparent discrepancy between the young ages 
inferred by the spectra compared to the old ages at which the majority of stars must have been formed.

Future work includes a similar study of other nearby ellipticals and also lenticular galaxies, 
searching for individual UV-bright young stars and open star clusters, to not only place tight constraints on 
their low-level star formation rates and histories, but to also address the role of environment and 
galaxy properties on star formation. 

\acknowledgments This work has made use of {\it BaSTI} web tools. We thank Adriano Pietrinferni
for providing stellar evolutionary tracks for the relevant WFC3 UVIS filters that were not publicly 
available. We also thank Jeremy Bailin, Eric Bell, and Sally Oey for providing valuable comments and 
assistance. Support for program \# 11583 was provided by NASA through a grant from the Space Telscope
Science Institute, which is operated by the Association of Universities for Research in Astronomy, Inc,. 
under NASA contract NAS 5-26555. The National Radio Astronomy Observatory is a facility of the 
National Science Foundation operated under cooperative agreement by Associated Universities, Inc.

\bibliography{ms.bib}

\end{document}